\documentclass[showpacs,prd,aps]{revtex4}

\pdfoutput=1

\usepackage{hyperref}
\usepackage{amsmath}
\usepackage{amssymb}
\usepackage{latexsym}
\usepackage{amsfonts}
\usepackage{epsfig}
\usepackage{psfrag}
\usepackage{graphicx}

\def\hhref#1{\href{http://arxiv.org/abs/#1}{arXiv:#1}} 

\usepackage{color}

\newcommand{\bea}{\begin{eqnarray}}
\newcommand{\ea}{\end{eqnarray}}
\newcommand{\eea}{\end{eqnarray}}

\begin{document}

\title{Interference Effects in Schwinger Vacuum Pair Production for Time-Dependent Laser Pulses}

\author{Cesim~K.~Dumlu and Gerald~V.~Dunne}

\affiliation{Department of Physics, University of Connecticut,
Storrs CT 06269-3046, USA}

\begin{abstract}
We present simple new approximate formulas, for both scalar and spinor QED, for the number of particles produced from vacuum by a time dependent electric field, incorporating the interference effects that arise from an arbitrary number of distinct semiclassical turning points. Such interference effects are important when the temporal profile of the laser pulse has subcycle structure. We show how the resulting semiclassical intuition may be used to guide the design of temporal profiles that enhance the momentum spectrum due to interference effects. The result is easy to implement and generally applicable to time-dependent tunneling problems, such as  appear in many other contexts in particle and nuclear physics, condensed matter physics, atomic physics, chemical physics, and gravitational physics.
\end{abstract}



\maketitle

\section{Introduction}

The Heisenberg-Schwinger effect is the non-perturbative production of electron-positron pairs when an external electric field is applied to  the quantum electrodynamical (QED) vacuum \cite{he,schwinger,greiner,dittrichgies,Ringwald:2001ib,gvd,Ruffini:2009hg}. It was one of the first non-trivial predictions of QED, but the effect is so weak that it has not yet been directly observed.
However,  new experimental developments in ultra-high intensity lasers \cite{tajima,eli} may soon bring us to the verge of this extreme ultra-relativistic regime \cite{mourou}. This experimental progress has renewed theoretical interest \cite{dunne-eli}, and recent results suggests that the effect may become observable in the $10^{25} -10^{26} W/cm^2$ intensity range, three or four orders of magnitude below the "Schwinger limit" of $4\times 10^{29} W/cm^2$, which comes from an estimate based on a constant electric field.  New theoretical ideas involve combining multiple copies of identical pulses \cite{Bulanov:2004de,Bulanov:2010ei}, and also shaping pulses in special ways using the "dynamically assisted Schwinger mechanism" \cite{Schutzhold:2008pz}, in which a  superposition of two time-dependent pulses, one strong but slow, and the other weak but fast, can lead to a significant enhancement of the tunneling process associated with the Schwinger effect. An explicit experimental realization  has been proposed \cite{DiPiazza:2009py} that suggests an observable rate of particle production. A closely related theoretical idea is that of a "catalyzed Schwinger mechanism" \cite{Dunne:2009gi}, which  can also be viewed as photon-stimulated pair-production \cite{Monin:2009aj}, realizing the more general mechanism of an induced metastable decay process \cite{voloshin}. The importance of cascading effects has been emphasized in \cite{Bell:2008zzb}. These, and other theoretical analyses of more realistic laser fields, such as plane waves of finite extent \cite{Heinzl:2010vg}, show that the precise form of the laser field can have a significant effect on the resulting pair production yield and momentum distribution. The strong sensitivity is not so surprising since it is a non-perturbative effect, but this makes it correspondingly difficult to do precise computations.

In the quantum field theoretic approach \cite{schwinger}, the theoretical problem is to compute the non-perturbative imaginary part of the "effective action", $\Gamma[A]=\hbar \, \ln\, \det\left[ i D\hskip -7pt /  -m\right]
$, where the Dirac operator, $D\hskip -7pt / \equiv \gamma^\mu (\partial_\mu-i\frac{e}{\hbar c}A_\mu)$, defines the coupling between electrons and the applied (classical) electromagnetic field $A_\mu$ that represents the field produced by the laser pulse. However, computationally we are currently limited to one-dimensional fields such as time dependent electric fields $E(t)$, in which case the problem can be more conveniently expressed as a "quantum mechanical" scattering problem, invoking Feynman's picture of anti-particles as particles traveling backward in time \cite{feynman-positron}. This requires the computation of a reflection probability for an over-the-barrier scattering problem, that can be done numerically or using WKB \cite{brezin,popov,popov-pulse,gavrilov,kimpage}, or in the quantum kinetic approach \cite{kluger,Rau:1995ea,schmidt,Blaschke:2008wf}. The WKB approach  is based on a relativistic extension of Keldysh's seminal work for atomic ionization in time-dependent electric fields \cite{keldysh}. Recently it has become clear that this WKB analysis must be extended to incorporate interference effects when the temporal profile of the laser pulse has sub-cycle structure.

For example, it has been shown \cite{Hebenstreit:2009km} that the momentum distribution of the produced particles is extremely sensitive to the "carrier phase" of a laser pulse, the phase offset between the pulse envelope and its oscillatory function. Moreover, this sensitivity reveals a distinct difference between spinor and scalar QED, which are conventionally treated on an equal footing at leading non-perturbative order. The oscillatory behavior of the momentum distribution is an interference effect, and can be quantitatively explained by incorporating the interference between different semiclassical saddle points. This was done in \cite{dd1} for the case of two distinct saddle points, using the phase integral method and the Stokes phenomenon. In this paper, Here we use another even simpler method, based on the Riccati form of the scattering problem,  and present new results for the case of an arbitrary number of distinct saddle points.

Physically, such interference phenomena are familiar from strong-field atomic and molecular physics, discussed long ago in the theory of atomic ionization \cite{terentev,popov-atomic-interference,popov-review}, and observed experimentally in photoionization spectra \cite{dalibard,paulus}. These ideas have even led to the proposal for an all-optical double-slit experiment in the time domain, using vacuum polarization effects \cite{Hebenstreit:2009km,king}.
They also appear naturally in any time-dependent tunneling effect, such as the Landau-Zener effect or other condensed matter systems \cite{oka}, chemical physics \cite{miller}, as well as gravitational \cite{padman,Kim:2010xm} and particle physics \cite{dima}.

In Section II we briefly review the scattering formalism for the QED pair production effect. In Section III we recall the numerical approach, and present our approximate expressions for the particle number. Sections IV and V contain explicit examples of particular temporal profiles for the electric field $E(t)$ that illustrate various features of the interference phenomena, and the final Section contains our conclusions.

\section{Scattering formalism}

In this Section we recall briefly  the scattering formalism of the pair production problem for both scalar and spinor QED, as we wish to compare the two cases in subsequent sections. For a linearly polarized electric field $\vec{E}=(0, 0, E(t))$ that is time dependent and pointing in the $x^3$ direction, we choose a vector potential $\vec{A}=(0, 0, A(t))$, with $E(t)=-\dot{A}(t)$. For such a field, spatial momentum is a good quantum number for the produced particles, so we can decompose the quantum field operators in terms of spatial momenta. For both scalar and spinor QED, the number of particles produced in each momentum mode can be expressed in terms of the reflection coefficient for an effective Schro\"dinger-like scattering problem. Physically, this is due to Feynman's interpretation of antiparticles as particles propagating backwards in time \cite{feynman-positron}, and has been used as a basic tool in the WKB analysis of the particle production problem \cite{brezin,popov,popov-pulse,kimpage}. The point of this current paper is to extend such semiclassical analyses to incorporate interference effects due to multiple saddle points, as this phenomenon naturally occurs for time dependent electric fields with sub-cycle structure, as is the case for more realistic representations of intense laser pulses.

\subsection{Scalar QED}

We decompose the scalar field operator as
\begin{eqnarray}
&&\Phi(\vec{x},t)=\int d^3k\,  e^{i\vec{k}\cdot \vec{x}} \left(\phi_{\bf{k}}(t)a_{\bf{k}}+\phi^{*}_{\bf{k}}(t)b^{\dagger}_{-\bf{k}}\right)
\label{modessca}
\end{eqnarray}
where $a_{\bf{k}}$ and $b^{\dagger}_{-\bf{k}}$ satisfy standard bosonic commutation relations, for each mode ${\bf k}$. The Klein-Gordon equation for $\Phi(\vec{x}, t)$ translates into the following  equation for the mode functions $\phi_{\bf{k}}(t)$:
\begin{eqnarray}
 \ddot{\phi}_{\bf{k}}(t) + Q^2_{\bf{k}}(t)\phi_{\bf{k}}(t)=0
\label{se}
\end{eqnarray}
where we define
\begin{eqnarray}
 Q^2_{\bf{k}}(t)= m^2 + k_{\perp}^2 + (k_{\parallel}-q A(t))^2
\label{q}
\end{eqnarray}
Equation (\ref{se}) has the form of a Schr\"odinger-like equation in the variable $t$
\begin{eqnarray}
-\ddot{\phi}_{\bf{k}}(t) -  (k_{\parallel}-q A(t))^2 \phi_{\bf{k}}(t)&=&( m^2 + k_{\perp}^2 ) \phi_{\bf{k}}(t)
\label{se2}
\end{eqnarray}
with "potential" $V(t)=-  (k_{\parallel}-q A(t))^2$, and "energy" $( m^2 + k_{\perp}^2 )$.
We implement the  Bogoliubov transformation by defining $\alpha_{\bf{k}}(t)$ and $\beta_{\bf{k}}(t)$ as follows:
\begin{eqnarray}
\phi_{\bf{k}}(t)&=&\frac{\alpha_{\bf{k}}(t)}{\sqrt{2 Q_{\bf{k}}(t)}}e^{-i\int^t Q_{\bf{k}}}+\frac{\beta_{\bf{k}}(t)}{\sqrt{2 Q_{\bf{k}}(t)}}e^{i\int^t Q_{\bf{k}}}
\nonumber\\
\dot{\phi}_{\bf{k}}(t)&=&-iQ_{\bf{k}}(t)\left(\frac{\alpha_{\bf{k}}(t)}{\sqrt{2 Q_{\bf{k}}(t)}}e^{-i\int^t Q_{\bf{k}}}-\frac{\beta_{\bf{k}}(t)}{\sqrt{2 Q_{\bf{k}}(t)}}e^{i\int^t Q_{\bf{k}}}\right)
\label{scalar-mode}
\end{eqnarray}
For each mode, the Bogoliubov coefficients, ${\alpha}_{\bf{k}}$ and ${\beta}_{\bf{k}}$, satisfy the first-order coupled equations:
\begin{eqnarray}
\dot{\alpha}_{\bf{k}}(t)&=&\frac{\dot{Q}_{\bf{k}}(t)}{2Q_{\bf{k}}(t)}\beta_{\bf{k}}(t)e^{2i\int^t Q_{\bf{k}}}\\
\dot{\beta}_{\bf{k}}(t)&=&\frac{\dot{Q}_{\bf{k}}(t)}{2Q_{\bf{k}}(t)}\alpha_{\bf{k}}(t)e^{-2i\int^t Q_{\bf{k}}}
\label{abdot}
\end{eqnarray}
This Bogoliubov transformation implements a change from the time-independent basis of creation and annihilation operators, $a_{\bf k}$ and $b_{-{\bf k}}^\dagger$,  to a time-dependent basis of creation and annihilation operators, $\tilde{a}_{\bf k}(t)$ and $\tilde{b}_{-{\bf k}}^\dagger(t)$, via the linear transformation:
\begin{eqnarray}
\begin{pmatrix}
\tilde{a}_{\bf k}(t)\cr
\tilde{b}_{-{\bf k}}^\dagger(t)
\end{pmatrix}
=\begin{pmatrix}
\alpha_{\bf k} & \beta_{\bf k}^*\cr
\beta_{\bf k} & \alpha_{\bf k}^*
\end{pmatrix}
\begin{pmatrix}
a_{\bf k}\cr
b_{-{\bf k}}
\end{pmatrix}
\end{eqnarray}
The bosonic commutation relations are preserved by the unitarity condition: $|\alpha_{\bf{k}}(t)|^2 -|\beta_{\bf{k}}(t)|^2=1$. The number of pairs produced in the momentum mode ${\bf k}$, from vacuum,  is given in terms of the modulus of the coefficient $\beta_{\bf k}$ at $t=+\infty$:
 \begin{eqnarray}
N_{\bf k}=|\beta_{\bf k}(t=+\infty)|^2
\label{n-scalar}
\end{eqnarray}
The relation to quantum mechanical scattering arises because we can express $N_{\bf k}$ in terms of the reflection probability, $|R_{\bf k}|^2=\left| \frac{\beta_{\bf k}(t)}{\alpha_{\bf k}(t)}\right |^2_{t=+\infty}$,  for the effective "Schr\"odinger" problem (\ref{se}):
 \begin{eqnarray}
N_{\bf k}
=\frac{|R_{\bf k}|^2}{1-| R_{\bf k}|^2}
\label{nr-scalar}
\end{eqnarray}
Recall from (\ref{se2}) that this describes the situation of over-the-barrier scattering, so the reflection probability is exponentially small, and so we can often make the approximation: $N_{\bf k}
\approx |R_{\bf k}|^2$.

\subsection{Spinor QED}

An analogous mode decomposition exists for spinor QED. We expand the spinor field operator  $\Psi(\vec{x}, t)$ as:
\begin{eqnarray}
\Psi(\vec{x},t)=\sum_{s}\int d^3k \, e^{i\vec{k}\cdot\vec{x}}\left(u_{\bf{k},\bf{s}}(t)\, a_{\bf{k},\bf{s}}+
v_{-\bf{k},\bf{s}}(t)\, b^{\dagger}_{-\bf{k},\bf{s}}\right)
\label{fermionmodes}
\end{eqnarray}
where $a_{\bf{k}}$ and $b^{\dagger}_{-\bf{k}}$ satisfy standard fermionic anti-commutation relations, for each mode ${\bf k}$, and the sum is over helicity $s=\pm 1$.
In a suitable Dirac matrix basis, the time dependent spinors, $u_{\bf{k},\bf{s}}(t)$ and $v_{\bf{k},\bf{s}}(t)$, can be written in terms of a single complex function $\psi_{\bf k}(t)$ that satisfies the Schr\"odinger-like equation:
\begin{eqnarray}
\ddot{\psi}_{\bf{k}}(t) + \left(Q^2_{\bf{k}}(t) + i\dot{k}_{\parallel}(t)\right)\psi_{\bf{k}}(t)=0
\label{fermionschr}
\end{eqnarray}
We implement the  Bogoliubov transformation by defining $\alpha_{\bf{k}}(t)$ and $\beta_{\bf{k}}(t)$ as follows:
\begin{eqnarray}
&&\psi_{\bf{k}}(t)=\frac{\alpha_{\bf{k}}(t)}{\sqrt{2 Q_{\bf{k}}(t)(Q_{\bf{k}}(t)-k_{\parallel}(t))}}e^{-i\int^t Q_{\bf{k}}}+\frac{\beta_{\bf{k}}(t)}{\sqrt{2 Q_{\bf{k}}(t)(Q_{\bf{k}}(t)+k_{\parallel}(t))}}e^{i\int^t Q_{\bf{k}}}
\\\nonumber
&&\dot{\psi}_{\bf{k}}(t)=-iQ_{\bf{k}}(t)\left(\frac{\alpha_{\bf{k}}(t)}{\sqrt{2 Q_{\bf{k}}(t)(Q_{\bf{k}}(t)-k_{\parallel}(t))}}e^{-i\int^t Q_{\bf{k}}}-\frac{\beta_{\bf{k}}(t)}{\sqrt{2 Q_{\bf{k}}(t)(Q_{\bf{k}}(t)+k_{\parallel}(t))}}e^{i\int^t Q_{\bf{k}}}\right)
\label{spinor-mode}
\end{eqnarray}
For each mode, the Bogoliubov coefficients satisfy the first-order coupled equations:
\begin{eqnarray}
\dot{\alpha}_{\bf{k}}(t)&=&\frac{\dot{k}_{\parallel}(t)\epsilon_{\perp}}{2Q^2_{\bf{k}}(t)}\beta_{\bf{k}}(t)e^{2i\int^t Q_{\bf{k}}}\\\dot{\beta}_{\bf{k}}(t)&=&-\frac{\dot{k}_{\parallel}(t)\epsilon_{\perp}}{2Q^2_{\bf{k}}(t)}\alpha_{\bf{k}}(t)e^{-2i\int^t Q_{\bf{k}}}.
\label{abdot2}
\end{eqnarray}
where $\epsilon_\perp^2\equiv m^2+k_\perp^2$.
This Bogoliubov transformation implements a change from the time-independent basis of creation and annihilation operators to a time-dependent basis of creation and annihilation operators, with the unitarity condition:
$|\alpha_{\bf{k}}(t)|^2+|\beta_{\bf{k}}(t)|^2=1$. Note the opposite sign from the scalar QED case. The number of pairs produced in the momentum mode ${\bf k}$, from vacuum, is given in terms of the modulus of the coefficient $\beta_{\bf k}$ at $t=+\infty$:
 \begin{eqnarray}
N_{\bf k}&=&|\beta_{\bf k}(t=+\infty)|^2\nonumber\\
&=&\frac{|R_{\bf k}|^2}{1+|R_{\bf k}|^2}
\label{nr-spinor}
\end{eqnarray}
where $|R_{\bf k}|^2$ is defined as the reflection probability, $|R_{\bf k}|^2=\left| \frac{\beta_{\bf k}(t)}{\alpha_{\bf k}(t)}\right |^2_{t=+\infty}$. Again, the reflection is typically very small, so we often make the approximation:
$N_{\bf k}\approx |R_{\bf k}|^2$.

\section{Scattering Formalism: Numerical and Semiclassical approaches}

\subsection{Scalar QED: numerical computation}

It is straightforward to convert the Schr\"odinger-like scattering problem (\ref{se2}) into a Riccati equation \cite{popov} that is suitable for simple numerical evaluation. (We now suppress the momentum mode label, ${\bf k}$, since all modes decouple, and so can be treated separately.)
From  equations (\ref{abdot}) it is clear that the reflection amplitude, $R=\beta/\alpha$, evolves with time as:
\begin{eqnarray}
\dot{R}&=&\frac{\alpha\,\dot{\beta}-\beta\,\dot{\alpha}}{\alpha^2}\nonumber\\
&=&\frac{\dot{Q}}{2Q}\left(e^{-2i\int^t Q}-R^2e^{2i\int^t Q}\right)
\label{rdot1}
\end{eqnarray}
This Riccati equation is trivial to integrate numerically, for a given $A(t)$ and longitudinal momentum $k_\parallel$, with the initial condition $R(-\infty)=0$, to obtain $R(\infty)$, whose magnitude squared gives the particle number (\ref{nr-scalar}). As discussed in \cite{Dumlu:2009rr}, this is completely equivalent to the quantum kinetic equation approach. We will use this numerical
formalism in order to obtain "exact" particle spectra, with which we can compare our semiclassical approximations.

\subsection{Spinor QED: numerical computation}

For spinor QED, the argument is very similar.
From the equations (\ref{abdot2}) it is clear that the reflection amplitude, $R=\beta/\alpha$, evolves with time as:
\begin{eqnarray}
\dot{R}&=&\frac{\alpha\,\dot{\beta}-\beta\,\dot{\alpha}}{\alpha^2}\nonumber\\
&=&-\frac{\dot{k_\parallel}\,\epsilon_\perp}{2Q^2}\left(e^{-2i\int^t Q}+R^2e^{2i\int^t Q}\right)
\label{rdot2}
\end{eqnarray}
Note the different signs  from the scalar case (\ref{rdot1}), and the different form of the function out the front. Again, it is simple to implement numerically, with the initial condition $R(-\infty)=0$, to obtain $R(+\infty)$, and hence the particle number from (\ref{nr-spinor}).

\subsection{Scalar QED: semiclassical approximation}

To motivate a semiclassical approximation to the Riccati equation (\ref{rdot1}), consider the fact that $R(t)$ is always small, and so neglect the nonlinear term on the right-hand-side \cite{visser}. Then we have simply,
\begin{eqnarray}
R(\infty)\approx\int_{-\infty}^\infty \frac{\dot{Q}}{2Q}\, e^{-2i\int^t_{-\infty} Q(t^\prime)\,dt^\prime}\, dt
\label{app1}
\end{eqnarray}
This integral is dominated by the contributions of the poles, where $Q=0$, which are the semiclassical turning points $t_p$.
In the neighborhood of such a turning point, change variables from $t$  to the "singulant" function
\begin{eqnarray}
\xi(t)=\int^t Q(t^\prime)\,dt^\prime
\end{eqnarray}
Now assume there is a first order zero of $Q^2(t)$ [a similar argument applies for other orders of poles], so that near the turning point, $Q\sim c\sqrt{t-t_p}$, and $\xi\sim\frac{2}{3}c\left(t-t_p\right)^{3/2}+\xi_p$. Then the approximate equation for $R(t)$ can be expressed as
\begin{eqnarray}
\frac{dR}{d\xi}\sim \frac{1}{6}\frac{1}{\xi-\xi_p}\,e^{-2i \xi}
\label{app2}
\end{eqnarray}
Therefore, each pole $\xi_p$ will contribute a term $R(\infty)\approx -\frac{\pi i}{3} e^{-2i \xi_p}$, where $\xi_p=\int_{-\infty}^{t_p}Q(t)\,dt$, and we have chosen to refer all the phase integrals to $t=-\infty$. In fact, this approximation does not give the correct prefactor. This prefactor problem was noted already in the seminal papers \cite{pokrovskii}, where it was resolved by comparison with soluble cases. In order to obtain the correct prefactor, we must consider also the nonlinear term in (\ref{rdot1}), and keep all the multiple-integral iteration terms. This procedure yields a prefactor of magnitude $1$ \cite{Berry:1972na,berry,kruskal,meyer}.
Using these results, we obtain an approximate expression for the reflection amplitude with a contribution from each turning point in the upper half complex plane:
\begin{eqnarray}
R(\infty)\approx \sum_{t_p} e^{-2 i \int_{-\infty}^{t_p}Q(t) \, dt}
\label{app3}
\end{eqnarray}
Integrals of $Q(t)$ along the real axis are real, while those along the imaginary direction are imaginary, so it is natural to split the exponents into phases and real parts. Let us define $s_p={\rm Re}(t_p)$ as the real part of a complex turning point $t_p$. Then we can separate out a common phase factor $e^{-2i\int_{-\infty}^{s_1}Q(t)\,dt}$ in the sum in (\ref{app3}), and write
\begin{eqnarray}
R(\infty)\approx e^{-2i\int_{-\infty}^{s_1}Q(t)\,dt} \left(\sum_{t_p} e^{-2i\,\theta_p}\, e^{-2 | \int_{s_p}^{t_p}Q(t) \, dt |}\right)
\label{app4}
\end{eqnarray}
where the phase, $\theta_p=\int_{s_1}^{s_p} Q(t)\,dt$, is the phase accumulated by integrating $Q(t)$ along the real axis between neighboring turning points. These phases incorporate the interference effect between distinct turning points and yield a simple expression for the reflection probability, when we take the modulus squared of the reflection amplitude in (\ref{app4}). Within this approximation, there is actually no distinction between the particle number $N$ in (\ref{nr-scalar}) and the reflection probability $|R|^2$, so we obtain the approximate expression:
\begin{eqnarray}
N^{\rm scalar}_{\bf k} \approx  \sum_{t_p}  e^{-2 K^{(p)}_{\bf k}}+\sum_{t_p\neq t_{p^\prime}} 2\cos\left(2\,\theta^{(p,p^\prime)}_{\bf k}\right) e^{- K_{\bf k}^{(p)}-K_{\bf k}^{(p^\prime)}}
\label{app-scalar}
\end{eqnarray}
where we have defined
\begin{eqnarray}
K_{\bf k}^{(p)}&\equiv & | \int_{t_p^*}^{t_p}Q_{\bf k}(t) \, dt |
\label{kt1}
\\
\theta^{(p,p^\prime)}_{\bf k}&\equiv & \int_{s_p}^{s_{p^\prime}} Q_{\bf k}(t)\,dt
\label{kt2}
\end{eqnarray}
We have restored the momentum label ${\bf k}$ to emphasize the fact that the answer depends on ${\bf k}$, because $Q_{\bf k}(t)$ depends on ${\bf k}$. It should of course be remembered that this means that the location of the turning points $t_p$ also depends on ${\bf k}$, and so do the interference terms $\theta^{(p,p^\prime)}_{\bf k}$. The first term in (\ref{app-scalar}) is the sum over the contributions of independent turning points, while the second sum characterizes the interference between different turning points. The dominant contributions are from turning points with the smallest values of $K_{\bf k}^{(p)}$, and interference effects are significant for pairs of turning points for which these integrals are comparable in magnitude. Loosely speaking, this often  corresponds to a rule of thumb that turning points closest to the real axis tend to dominate, and interference effcts are strongest between pairs of turning points that have comparable distance from the real axis.

For later use, we record the approximate expressions for one, two and three complex conjugate pairs of turning points. If a single turning point pair dominates, then we have the familiar textbook expression \cite{landau}:
\begin{eqnarray}
N^{\rm scalar}_{\bf k} \approx   e^{-2K_{\bf k}^{(p)}}
\label{single-scalar}
\end{eqnarray}
If there are two pairs of turning points, $(t_1, t_1^*)$ and $(t_2, t_2^*)$, with comparable real exponential factors $e^{-2K_{\bf k}^{(1)}}$ and $e^{-2K_{\bf k}^{(2)}}$,
then there is a single interference term
\begin{eqnarray}
N^{\rm scalar}_{\bf k} \approx e^{-2 K_{\bf k}^{(1)}}+e^{-2 K_{\bf k}^{(2)}}+2\cos\left(2\,\theta^{(1,2)}_{\bf k}\right) e^{- K_{\bf k}^{(1)}-K_{\bf k}^{(2)}}
\label{two-scalar}
\end{eqnarray}
where $\theta^{(1,2)}_{\bf k}=\int_{s_1}^{s_2} Q_{\bf k}(t) \, dt$. This is the case that was studied in \cite{dd1}. If there are three turning point pairs, $(t_1, t_1^*)$,  $(t_2, t_2^*)$, and $(t_3, t_3^*)$, each with comparable real exponential factors $e^{-2 K_{\bf k}^{(p)}}$, then there are three interference terms:
\begin{eqnarray}
N^{\rm scalar}_{\bf k} &\approx& e^{-2 K_{\bf k}^{(1)}}+e^{-2 K_{\bf k}^{(2)}}+e^{-2 K_{\bf k}^{(3)}}+2\cos\left(2\,\theta^{(1,2)}_{\bf k}\right) e^{- K_{\bf k}^{(1)}-K_{\bf k}^{(2)}}\nonumber\\
&&
+2\cos\left(2\,\theta^{(2,3)}_{\bf k}\right) e^{- K_{\bf k}^{(2)}-K_{\bf k}^{(3)}}
+2\cos\left(2\,\theta^{(1,3)}_{\bf k}\right) e^{- K_{\bf k}^{(1)}-K_{\bf k}^{(3)}}
\label{three-scalar}
\end{eqnarray}
where $\theta^{(1,2)}_{\bf k}=\int_{s_1}^{s_2} Q_{\bf k}(t) \, dt$, $\theta^{(2,3)}_{\bf k}=\int_{s_2}^{s_3} Q_{\bf k}(t) \, dt$, and $\theta^{(1,3)}_{\bf k}=\int_{s_1}^{s_3} Q_{\bf k}(t) \, dt$. The extension to more pairs of turning points is clear.

In the next Section we will illustrate these interference effects with explicit examples of electric fields that produce exactly one, two and three pairs of turning points. In the semiclassical regime, the expression (\ref{app-scalar}) is an excellent approximation, and describes the interference effects both qualitatively and quantitatively for a broad range of physical parameters.

\subsection{Spinor QED: semiclassical approximation}

For spinor QED, we can apply a similar argument to the Riccati equation (\ref{rdot2}). The difference is that the initial approximation, which is then iterated, yields a different function:
\begin{eqnarray}
R(\infty)\approx-\int_{-\infty}^\infty \frac{\dot{k_\parallel}\, \epsilon_\perp}{2Q^2}\, e^{-2i\int^t Q}\, dt
\label{app1-spinor}
\end{eqnarray}
In the vicinity of a turning point, we have
\begin{eqnarray}
- \frac{\dot{k_\parallel}\, \epsilon_\perp}{2Q^2}\sim \frac{\dot{Q}}{2Q}\, \frac{\epsilon_\perp}{k-A(t)}\sim \pm i \frac{\dot{Q}}{2Q}
\end{eqnarray}
with the sign depending on the branch. These signs alternate between successive turning points, so we obtain an extra (alternating sign) phase
\begin{eqnarray}
R(\infty)\approx \sum_{t_p} (-1)^p\, e^{i\, \pi/2}\, e^{-2 i \int_{-\infty}^{t_p}Q(t) \, dt}
\label{app3-spinor}
\end{eqnarray}
which leads to an approximate expression for the particle number for spinor QED:
\begin{eqnarray}
N^{\rm spinor}_{\bf k} \approx   \sum_{t_p}  e^{-2 K^{(p)}_{\bf k}}+\sum_{t_p\neq t_{p^\prime}} 2\cos\left(2\,\theta^{(p,p^\prime)}_{\bf k}\right)(-1)^{(p-p^\prime)}\, e^{- K_{\bf k}^{(p)}-K_{\bf k}^{(p^\prime)}}
\label{app-spinor}
\end{eqnarray}
where $K^{(p)}_{\bf k}$ and $\theta^{(p,p^\prime)}_{\bf k}$ are defined exactly as in (\ref{kt1}) and (\ref{kt2}).
The only difference from the scalar QED case lies in the signs of the interference terms. For example, if a single turning point pair dominates, because of a dominant real factor $e^{-2 | \int_{t_p^*}^{t_p}Q_{\bf k}(t) \, dt |}$, then there is no interference and we have just
\begin{eqnarray}
N^{\rm spinor}_{\bf k} \approx e^{-2K_{\bf k}^{(p)}}
\label{single-spinor}
\end{eqnarray}
which is the same as for scalar QED.
If there are two pairs of turning points, $(t_1, t_1^*)$ and $(t_2, t_2^*)$, with comparable real exponential factors
$e^{-2K_{\bf k}^{(1)}}$ and $e^{-2K_{\bf k}^{(2)}}$,
 then there is a single interference term, with the opposite sign from the scalar case:
\begin{eqnarray}
N^{\rm spinor}_{\bf k} \approx e^{-2 K_{\bf k}^{(1)}}+e^{-2 K_{\bf k}^{(2)}}-2\cos\left(2\,\theta^{(1,2)}_{\bf k}\right) e^{- K_{\bf k}^{(1)}-K_{\bf k}^{(2)}}
\label{two-spinor}
\end{eqnarray}
This is the case that was studied in \cite{dd1}. If there are three turning point pairs, $(t_1, t_1^*)$,  $(t_2, t_2^*)$, and $(t_3, t_3^*)$, each with comparable real exponential factors $e^{-2 K_{\bf k}^{(p)}}$, then there are three interference terms, with signs as follows:
\begin{eqnarray}
N^{\rm spinor}_{\bf k} &\approx& e^{-2 K_{\bf k}^{(1)}}+e^{-2 K_{\bf k}^{(2)}}+e^{-2 K_{\bf k}^{(3)}}-2\cos\left(2\,\theta^{(1,2)}_{\bf k}\right) e^{- K_{\bf k}^{(1)}-K_{\bf k}^{(2)}}\nonumber\\
&&
-2\cos\left(2\,\theta^{(2,3)}_{\bf k}\right) e^{- K_{\bf k}^{(2)}-K_{\bf k}^{(3)}}
+2\cos\left(2\,\theta^{(1,3)}_{\bf k}\right) e^{- K_{\bf k}^{(1)}-K_{\bf k}^{(3)}}
\label{three-spinor}
\end{eqnarray}
where $\theta^{(1,2)}_{\bf k}=\int_{s_1}^{s_2} Q_{\bf k}(t) \, dt$, $\theta^{(2,3)}_{\bf k}=\int_{s_2}^{s_3} Q_{\bf k}(t) \, dt$, and $\theta^{(1,3)}_{\bf k}=\int_{s_1}^{s_3} Q_{\bf k}(t) \, dt$. The extension to more pairs of turning points is clear.

\section{Illustrative Examples}

In this section, we compare our semiclassical approximations (\ref{app-scalar}) and (\ref{app-spinor}) that incorporate interference effects, with the (exact) numerical approach based on the Riccati equations (\ref{rdot1}) and (\ref{rdot2}), for both scalar and spinor QED. For this comparison, we have constructed electric fields such that the corresponding over-the-barrier scattering problem has precisely one, two and three pairs of complex conjugate turning points.

\subsection{One Pair of Turning Points}
\begin{figure}[htb]
\includegraphics[scale=0.7]{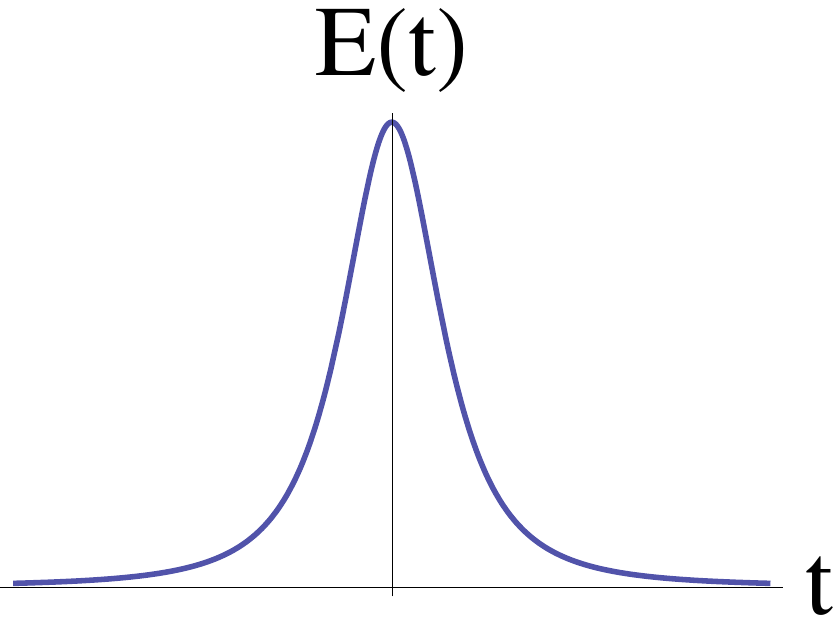}\qquad\qquad
\includegraphics[scale=0.7]{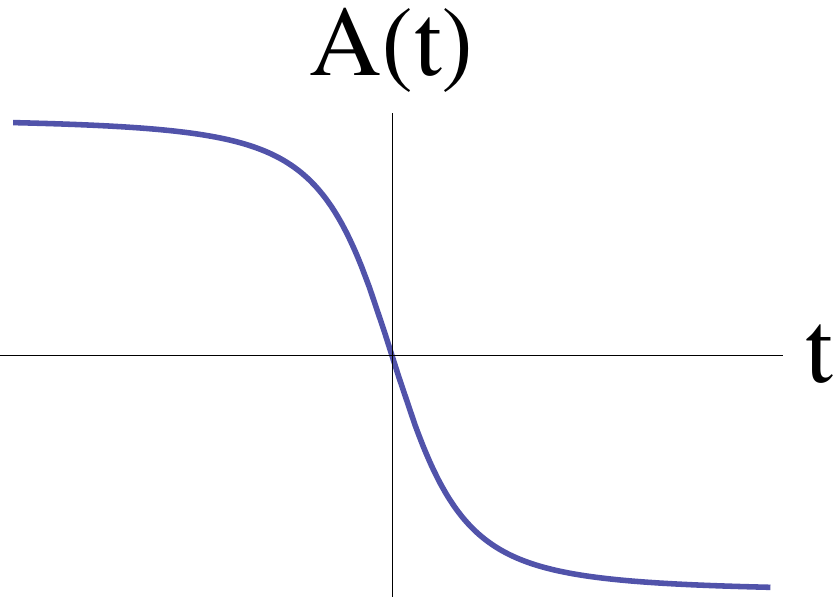}
\caption{The form of the electric field $E(t)$ in (\ref{e1}), and corresponding vector potential $A(t)$ in (\ref{a1}) for a single complex conjugate pair of turning points. $E(t)$ is an even function, while $A(t)$ is an odd function.}
\label{fig1}
\end{figure}

\begin{figure}[htb]
\includegraphics[scale=0.6]{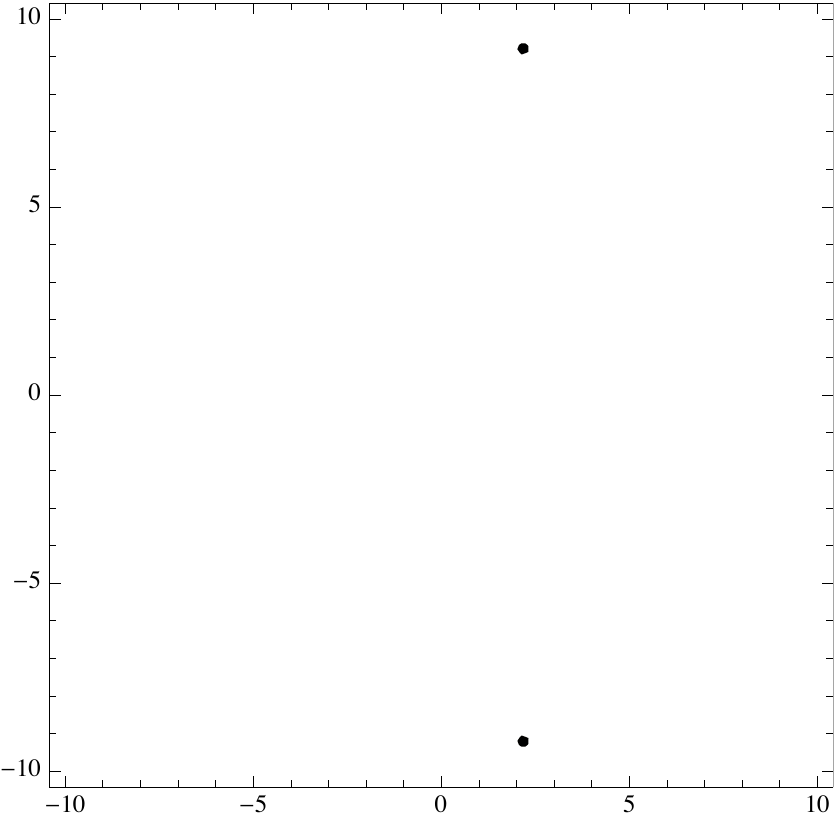}
\includegraphics[scale=0.6]{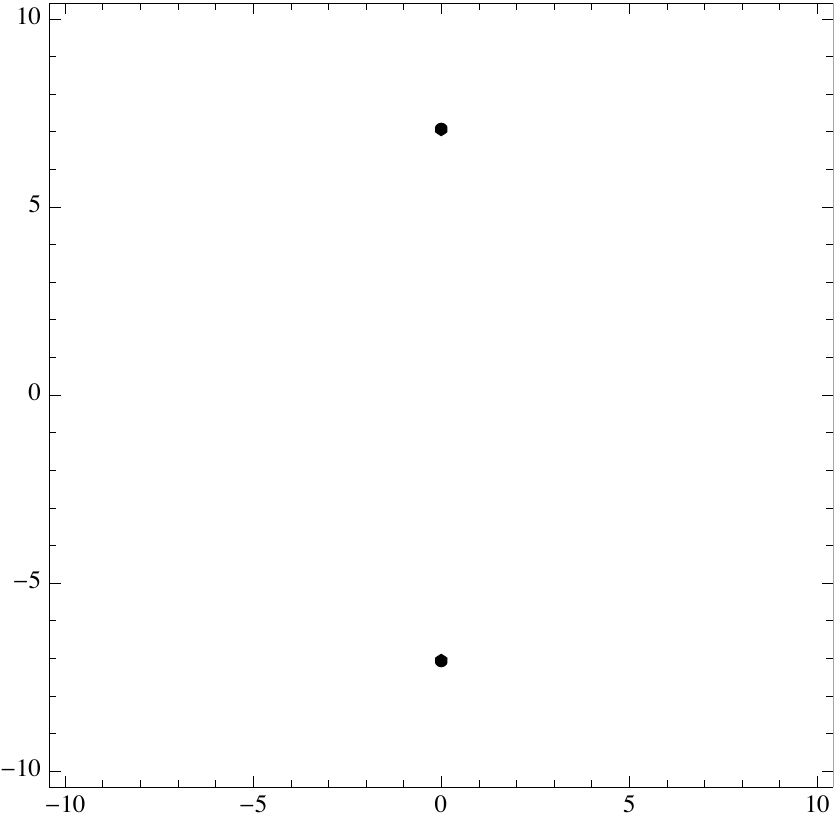}
\includegraphics[scale=0.6]{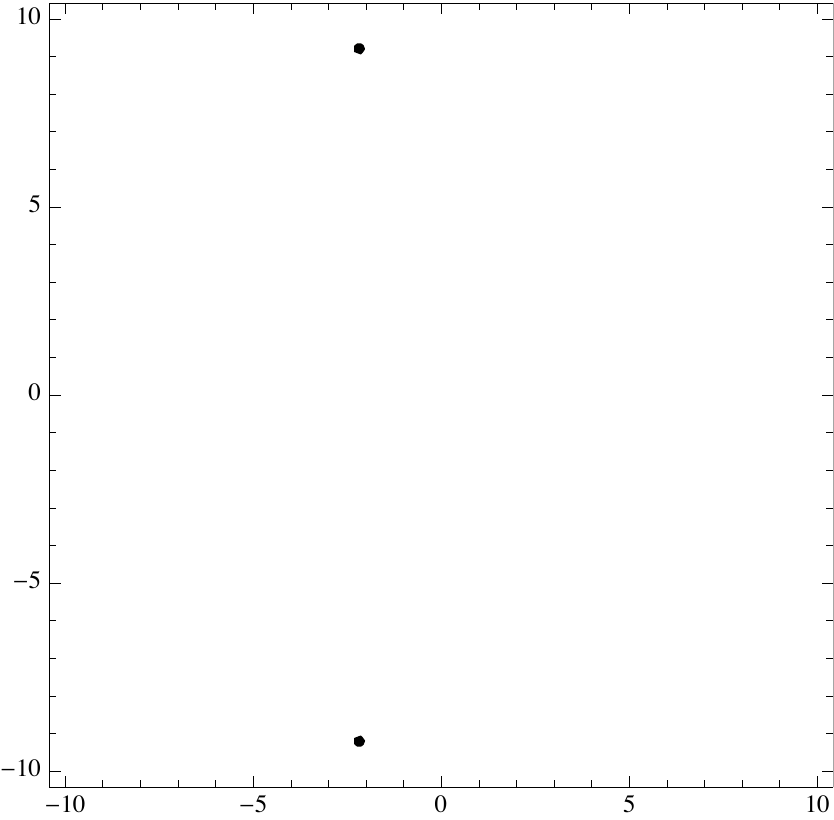}
\caption{The locations of the complex conjugate pair of turning points, in the complex $t$ plane, for three different values of longitudinal momentum. These plots are for the vector potential $A(t)$ in (\ref{a1}), with $E_0=0.1$ and $\omega=0.1$, for longitudinal momentum vales $k_\parallel=-1$ (left), $k_\parallel=0$ (center), and $k_\parallel=1$ (right), in units with $m=1$. Note that the turning points are closest to the real axis for $k_\parallel=0$.}
\label{fig2}
\end{figure}

An example of a gauge field with only single pair of turning points is the single-bump field
\begin{eqnarray}
E(t)=\frac{E_0}{\left(1 + \omega^2t^2\right)^{3/2}}
\label{e1}
\end{eqnarray}
where $E_0$ is the field strength amplitude, and $\omega$ is the inverse width, as shown in the left panel of Figure \ref{fig1}. The associated vector potential can be taken as
\begin{equation}
A(t)=-\frac{E_0\,t}{\sqrt{1+\omega^2t^2}}
\label{a1}
\end{equation}
This vector potential is plotted in the right panel of Figure \ref{fig1}. Note that $E(t)$ is an even function, while $A(t)$ is an odd function. This field has exactly one pair of complex conjugate turning points,  $(t_1(k), t_1^*(k))$, with
\begin{eqnarray}
t_1(k)=\frac{-(k-i)}{\sqrt{E_0^2 + \omega^2 +2i k \omega^2 - k^2 \omega^2}}
\end{eqnarray}
Note that as a function of the longitudinal momentum, $k$, the pair of turning points moves around in the complex plane, as shown in Figure \ref{fig2}, but remain a complex conjugate pair.

Figure 3 shows a comparison between the approximations (\ref{single-scalar}) and (\ref{single-spinor}) and the exact numerical results, for the particle number as a function of longitudinal momentum. The left plot is for scalar QED and the right plot is for spinor QED.
There is no oscillatory structure  in this momentum spectrum, as expected since there is no interference term for just a single pair of turning points. Thus, the spectra are the same for scalar and spinor QED, and there is good agreement between the approximate and exact results.
While the form of the electric  field (\ref{e1}) was chosen so that there is precisely one complex conjugate pair of turning points, similar behavior is obtained for other "single-bump" electric fields such as $E(t)=E_0\, {\rm sech}^2(\omega t)$, or $E(t)=E_0\exp(-\omega^2\,t^2)$, for which there is an infinite tower of turning points pairs, but only one pair [the one closest to the real axis] dominates, and the approximate expressions (\ref{single-scalar}) and (\ref{single-spinor}) again provide extremely accurate answers.
\begin{figure}[htb]
\begin{center}$
\begin{array}{cc}
\includegraphics[scale=0.75]{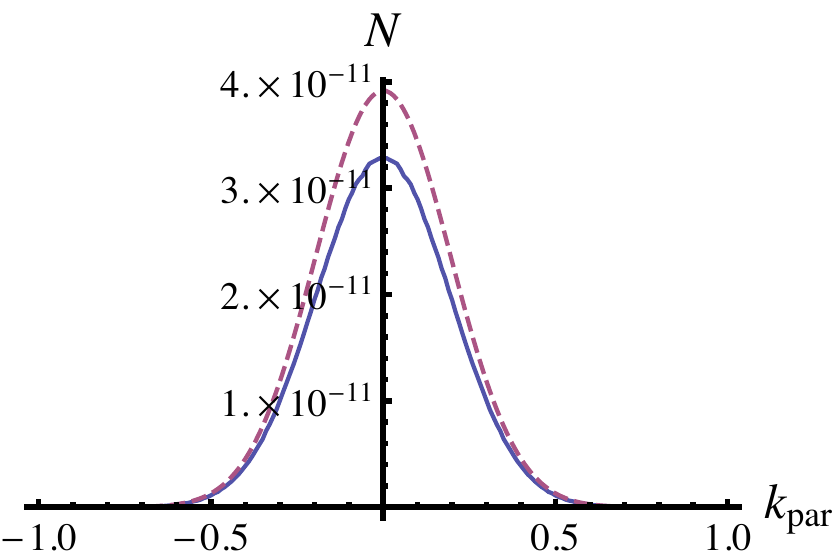}&
\includegraphics[scale=0.75]{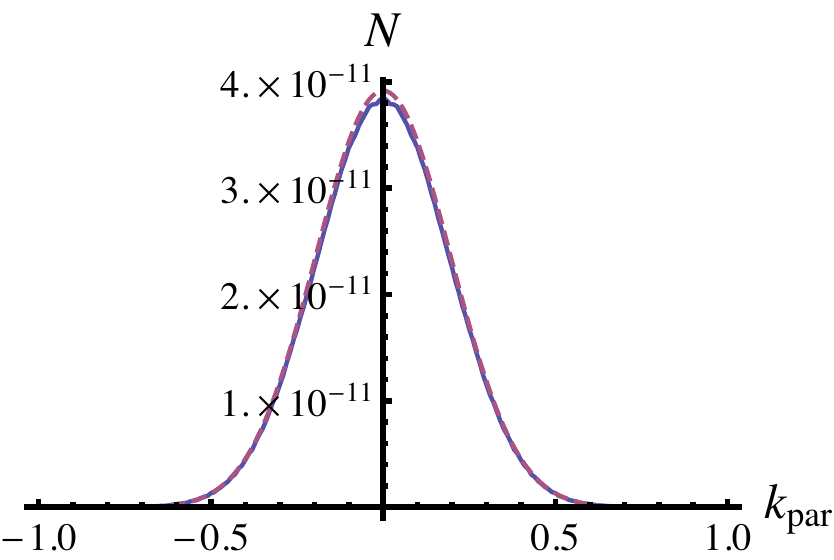}
\end{array}$
\end{center}
\caption{Scalar (left) and spinor (right) QED momentum spectra for vacuum pair production, as a function of longitudinal momentum,  for the electric field (\ref{e1}) that has one pair of turning points. The thick (blue) lines show the numerical calculation,  and the dashed (red) lines show the approximate expressions (\ref{single-scalar}) and (\ref{single-spinor}).  The field parameters were chosen as: $E_0=0.1$, and  $\omega=0.1$, in units with $m=1$. }
\label{fig3}
\end{figure}

\subsection{Two Pairs of Turning Points}
\begin{figure}[htb]
\includegraphics[scale=0.7]{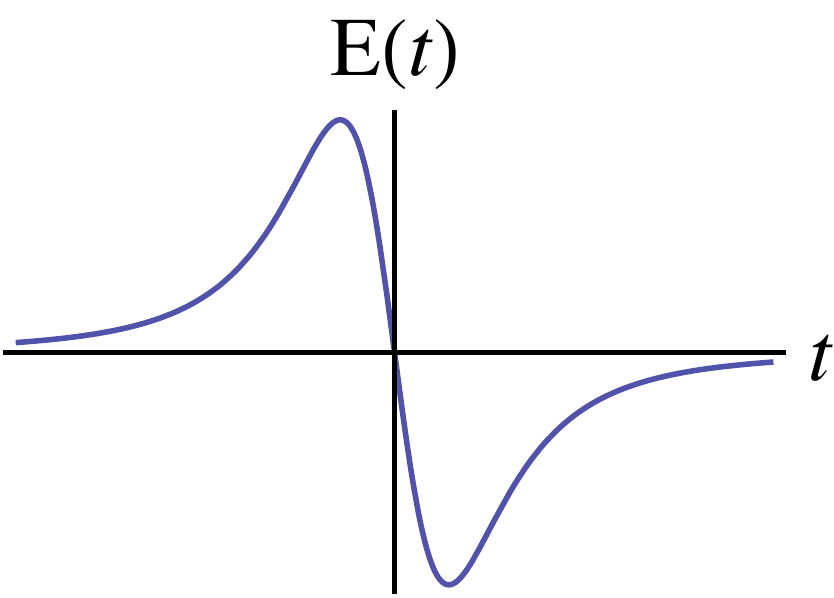}\qquad
\includegraphics[scale=0.7]{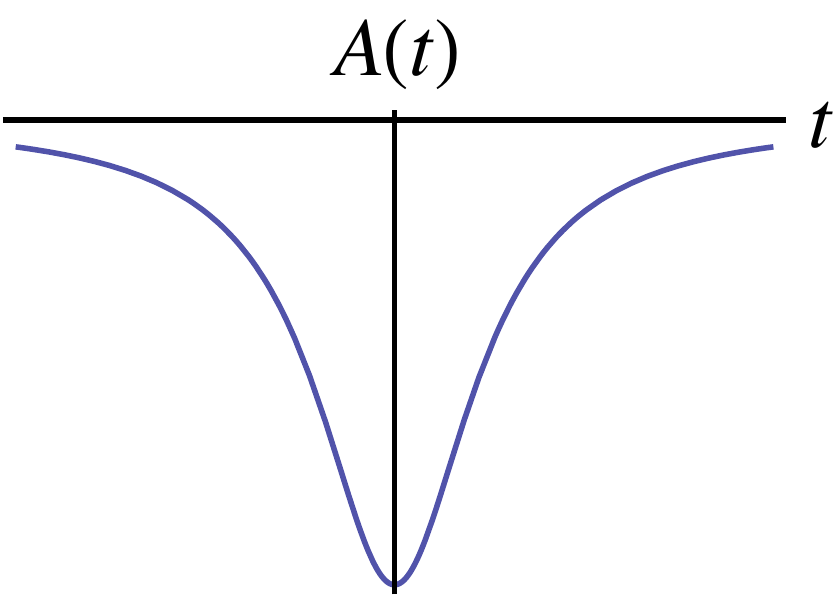}
\caption{The form of the electric field $E(t)$ in (\ref{e2}), and corresponding vector potential $A(t)$ in (\ref{a2}) for two complex conjugate pairs of turning points. $E(t)$ is an odd function, while $A(t)$ is an even function.}
\label{fig4}
\end{figure}

\begin{figure}[htb]
\includegraphics[scale=0.6]{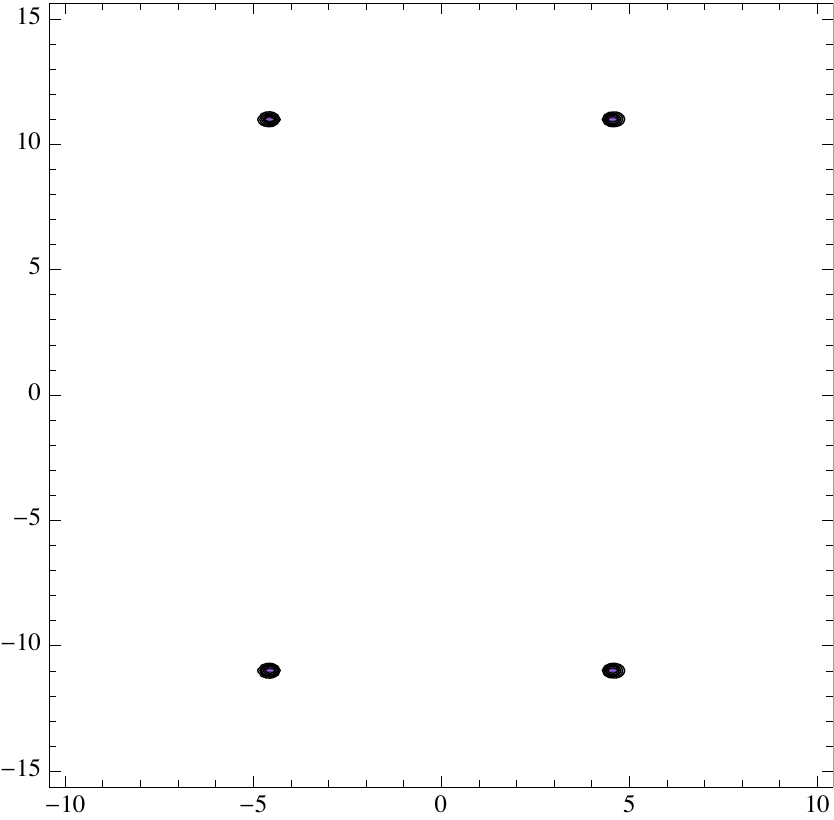}
\includegraphics[scale=0.6]{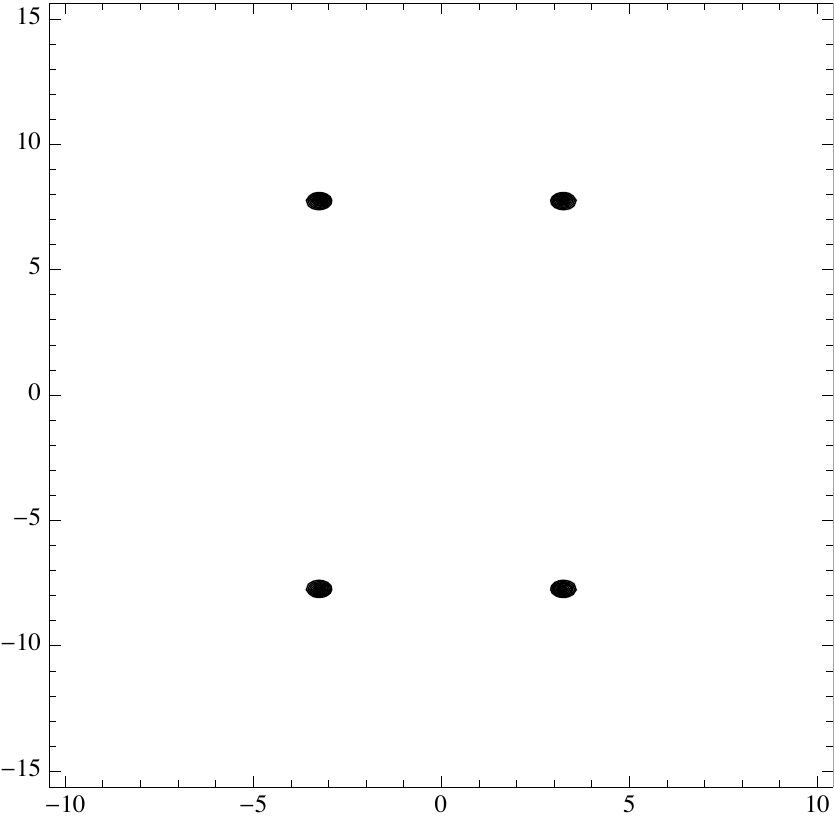}
\includegraphics[scale=0.6]{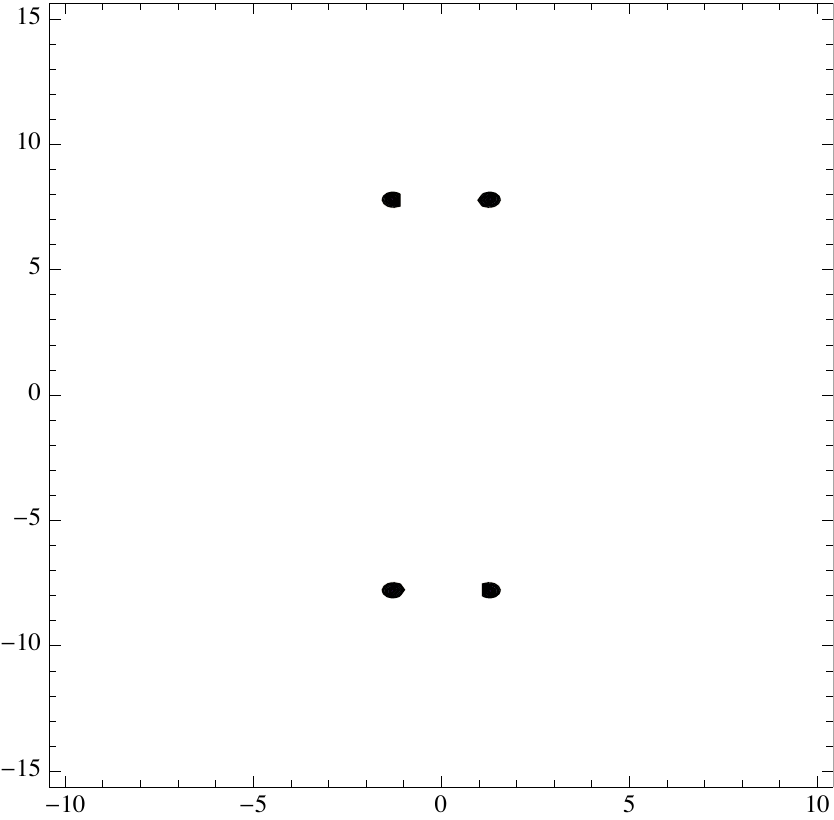}
\caption{The locations of the complex conjugate pair of turning points, in the complex $t$ plane, for three different values of longitudinal momentum. These plots are for the vector potential $A(t)$ in (\ref{a2}), with $E_0=0.1$ and $\omega=0.1$, for longitudinal momentum vales $k_\parallel=0$ (left), $k_\parallel=1$ (center), and $k_\parallel=2$ (right), in units with $m=1$. Note that the two turning points are always equidistant from the real axis, and note that they are closest to the real axis at a nonzero value of $k_\parallel$, which for these parameters is $k_\parallel\approx 1.2$.}
\label{fig5}
\end{figure}

\begin{figure}[htb]
\begin{center}$
\begin{array}{cc}
\includegraphics[scale=0.8]{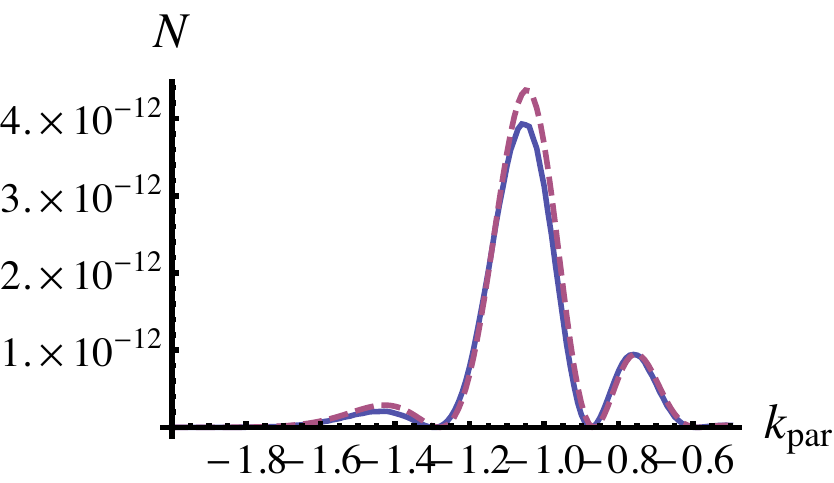}\qquad 
\includegraphics[scale=0.8]{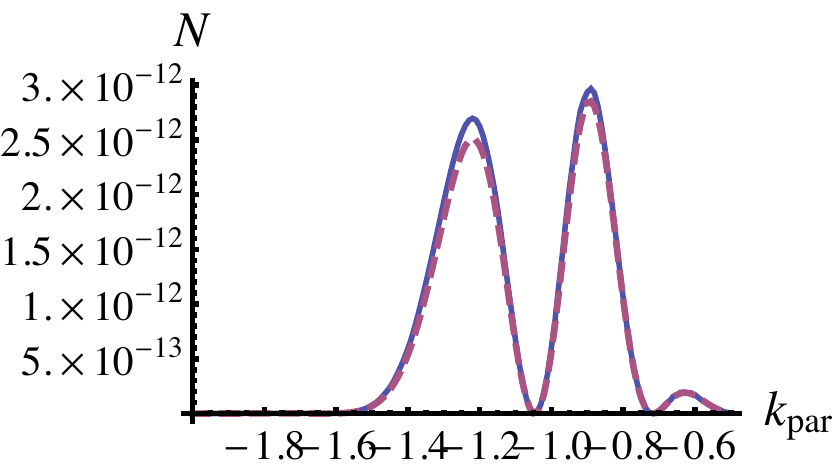}
\end{array}$
\end{center}
\caption{Scalar (left) and spinor (right) QED momentum spectra for vacuum pair production, as a function of longitudinal momentum,  for the electric field (\ref{e2}) that has two pairs of turning points. The thick (blue) lines show the numerical calculation,  and the dashed (red) lines show the approximate expressions (\ref{two-scalar}) and (\ref{two-spinor}).  The field parameters were chosen as: $E_0=0.1$, and $\omega=0.1$, in units with $m=1$.}
\label{fig6}
\end{figure}

To illustrate the effect of interference between pairs of turning points, we now consider an example of a vector potential leading to precisely two pairs of complex conjugate turning points. This field was considered already in \cite{dd1}, and here we give more details. Consider the electric field
\begin{eqnarray}
E(t)= -\frac{2 E_0 \omega t}{\left(1 + \omega^2t^2\right)^2}
\label{e2}
\end{eqnarray}
where $E_0$ is the field strength amplitude, and $\omega$ is the inverse width, as shown in the left panel of Figure \ref{fig4}. The associated vector potential can be taken as
\begin{equation}
A(t)=-\frac{E_0/\omega}{(1+\omega^2 \, t^2)}
\label{a2}
\end{equation}
This vector potential is plotted in the right panel of Figure \ref{fig4}. Note that $E(t)$ is an odd function, while $A(t)$ is an even function.
For this vector potential, there are  two complex conjugate pairs of turning points, $(t_1(k), t_1^*(k))$, and $(t_2(k), t_2^*(k))$, where:
\begin{eqnarray}
t_1(k)&=& \frac{\sqrt{-E_0-k \omega +i \omega }}{\omega^{3/2}\, \sqrt{k-i} } \\
t_2(k)&=&-\frac{\sqrt{-E_0-k \omega -i \omega}}{ \omega ^{3/2}\,\sqrt{k+i}}
\end{eqnarray}
These turning points are illustrated in Figure \ref{fig5}. An important difference from the case of a single pair of turning points shown in Figure \ref{fig2} is that now the point of closest approach of the turning points to the real axis occurs at a nonzero value of $k_\parallel$. This is reflected in the momentum spectrum for the two-pair case, shown in Figure 6, which is centered around a non-zero value of $k_\parallel$, while the  momentum spectrum for the single-pair case, shown in Figure 3, is centered around  $k_\parallel=0$. Also, observe that since the two pairs are equidistant from the real axis, we should expect strong interference effects between the two pairs of turning points, as indeed is seen in Figure \ref{fig6} for both scalar and spinor QED.

Figure \ref{fig6} shows a comparison between the approximations (\ref{two-scalar}) and (\ref{two-spinor}) and the exact numerical results, for the particle number as a function of longitudinal momentum. Note the oscillatory behavior of the spectrum, due to the interference terms. Also notice that the interference term has the opposite sign for scalar and spinor QED, as reflected in the exact momentum spectrum. The agreement between the exact numerical results [solid, blue lines] and the approximate semiclassical expressions [dashed, red lines] is extremely good, both qualitatively and quantitatively.

The form of the electric  field (\ref{e2}) was chosen so that there are  precisely two complex conjugate pairs of turning points. In fact, for other electric fields with temporal profile that is an odd function of $t$, as in  Figure \ref{fig4}, we find that there are two dominant pairs of turning points. For example, this occurs when $E(t)=E_0\,\omega \, t\, {\rm sech}^2(\omega t)$, or  $E(t)=E_0\,\omega\,t\, \exp(-\omega^2\,t^2)$, for which there is an infinite tower of turning points pairs, but only two pairs [those closest to the real axis] dominate, and the approximate expressions (\ref{two-scalar}) and (\ref{two-spinor}) again provide extremely accurate answers.

\subsection{Three Pairs of Turning Points}

\begin{figure}[htb]
\includegraphics[scale=0.7]{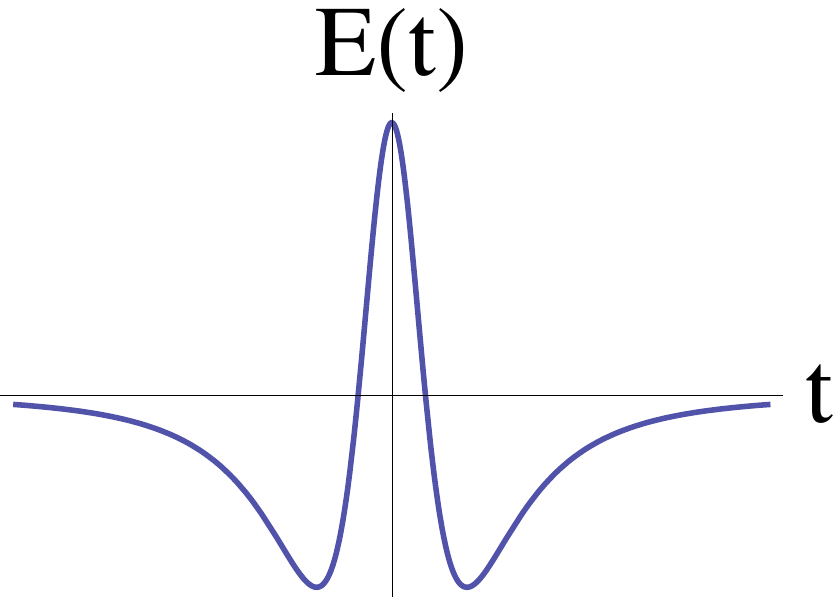}\qquad
\includegraphics[scale=0.7]{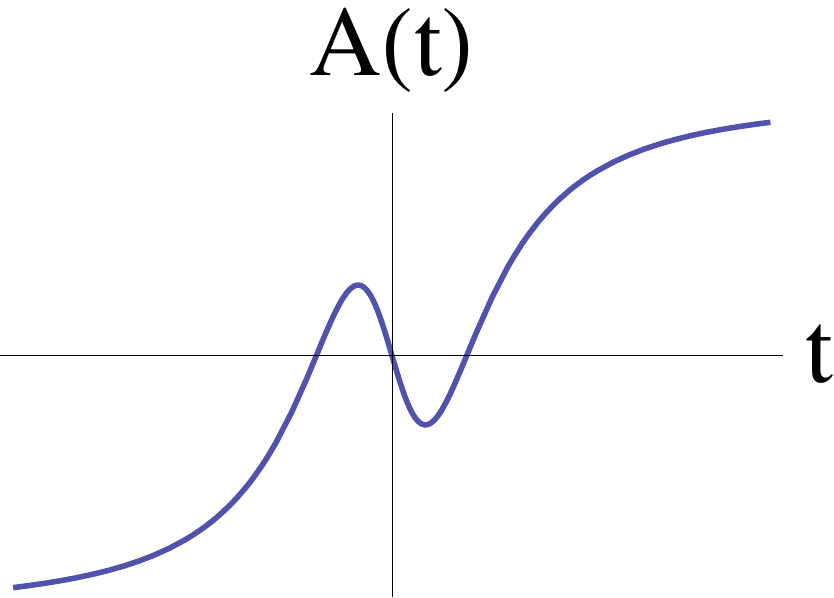}
\caption{The form of the electric field $E(t)$ in (\ref{e3}), and corresponding vector potential $A(t)$ in (\ref{a3}) for three complex conjugate pairs of turning points. $E(t)$ is an even function, while $A(t)$ is an odd function.}
\label{fig7}
\end{figure}

\begin{figure}[htb]
\includegraphics[scale=0.6]{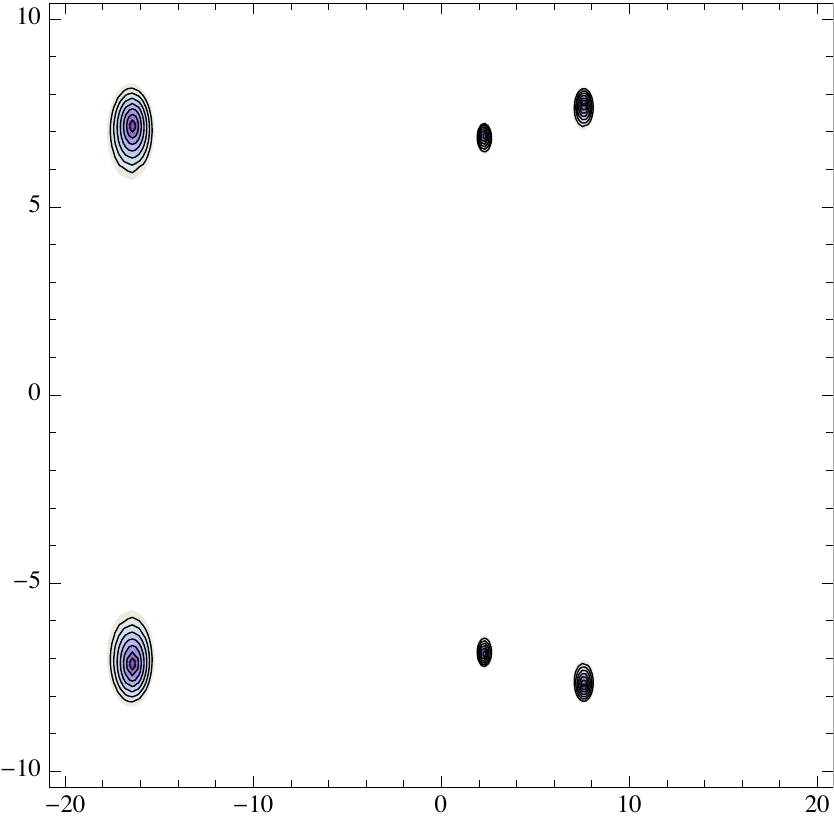}
\includegraphics[scale=0.6]{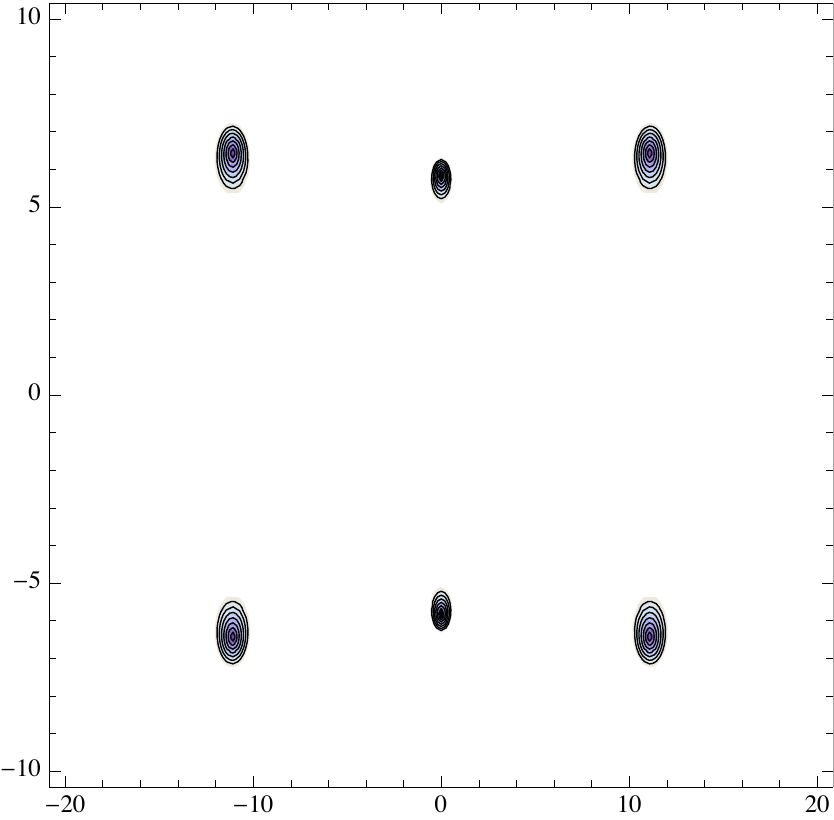}
\includegraphics[scale=0.6]{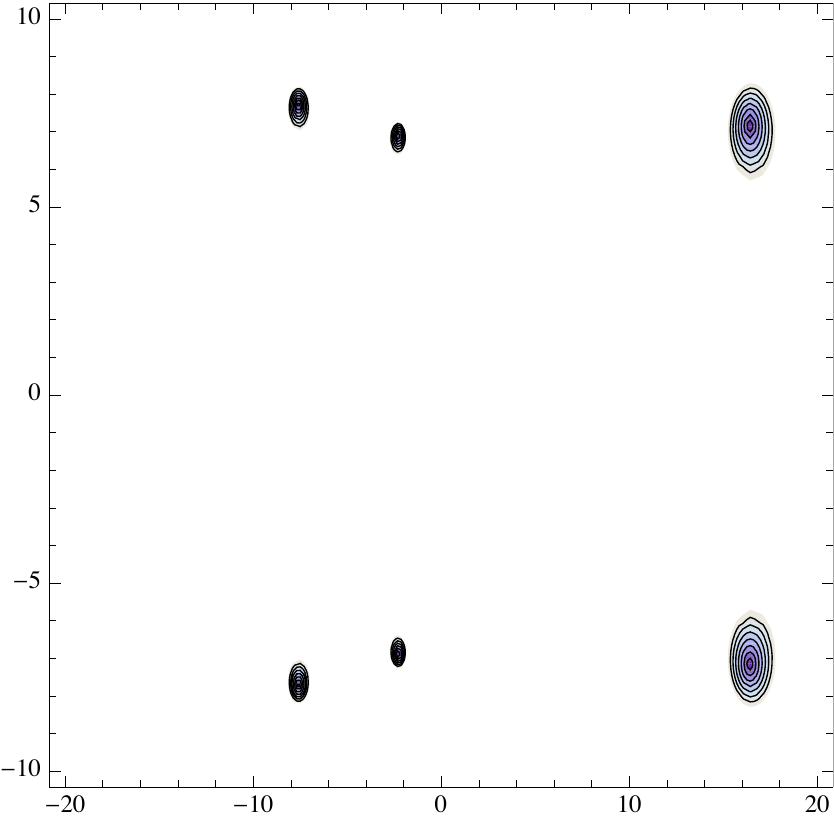}
\caption{The locations of the complex conjugate pair of turning points, in the complex $t$ plane, for three different values of longitudinal momentum. These plots are for the vector potential $A(t)$ in (\ref{a3}), with $E_0=0.1$, $\omega_1=0.1$, and $\omega_2=1/15$, for longitudinal momentum vales $k_\parallel=-1$ (left), $k_\parallel=0$ (center), and $k_\parallel=1$ (right), in units with $m=1$. Their distribution suggests we should expect significant interference effects, especially near $k_\parallel=0$.}
\label{fig8}
\end{figure}

\begin{figure}[htb]
\begin{center}$
\begin{array}{cc}
\includegraphics[scale=0.75]{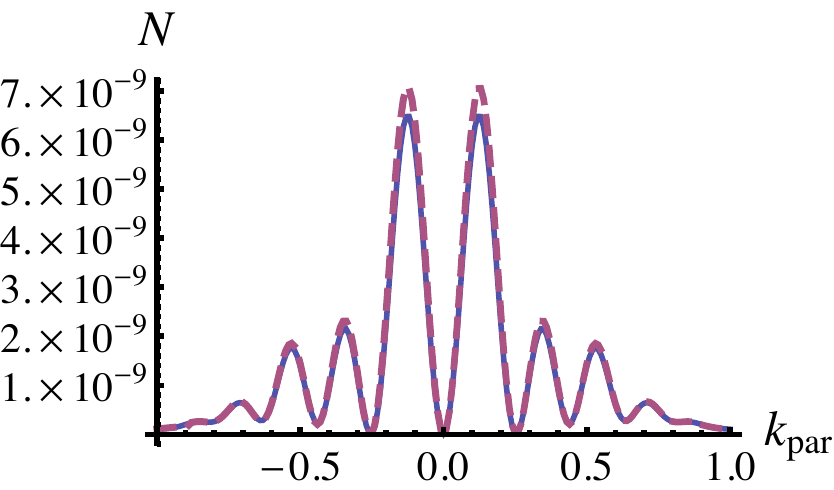}\qquad 
\includegraphics[scale=0.75]{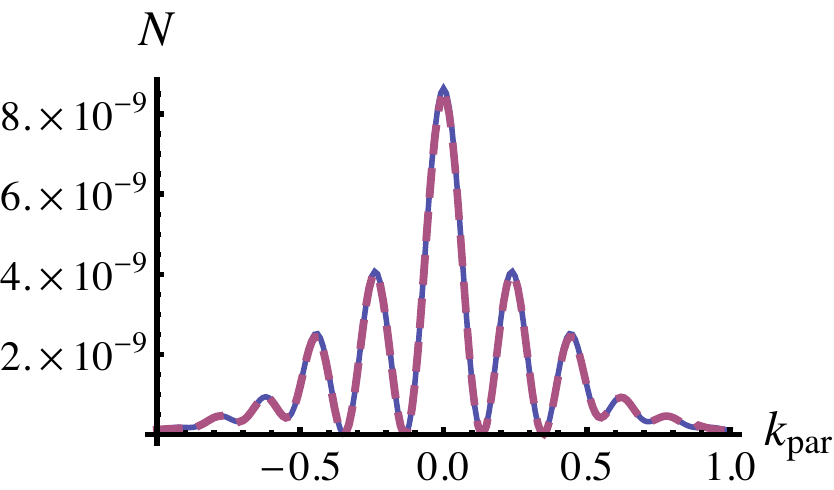}
\end{array}$
\end{center}
\caption{Scalar (left) and spinor (right) QED momentum spectra for vacuum pair production, as a function of longitudinal momentum,  for the electric field (\ref{e3}) that has three pairs of turning points. The thick (blue) lines show the numerical calculation,  and the dashed (red) lines show the approximate expressions (\ref{three-scalar}) and (\ref{three-spinor}).  The field parameters were chosen as: $E_0=0.1$, $\omega_1=0.1$, and $\omega_2=1/15$, in units with $m=1$.}
\label{fig9}
\end{figure}

To illustrate further the effect of interference between pairs of turning points, we now consider an example of a vector potential leading to precisely three pairs of complex conjugate turning points. This goes beyond the field considered already in \cite{dd1}, and permits us to verify the sign pattern of the interference terms in the spinor QED expression (\ref{three-spinor}). Consider the electric field
\begin{eqnarray}
E(t)= \frac{E_0 \left(1 -  \left(3 \omega_1^2 + 2 \omega_2^2\right)t^2\right)}{\left(1 + \omega_2^2t^2\right)^{5/2}}
\label{e3}
\end{eqnarray}
where $E_0$ is the field strength amplitude, and $\omega_1$ and $\omega_2$ represent two independent  inverse width scales. The form of this electric field is shown in the left panel of Figure \ref{fig7}. The associated vector potential can be taken as
\begin{equation}
A(t)=-\frac{E_0 t\left(1-\omega_1^2\,t^2\right)}{(1+\omega_2^2 \, t^2)^{3/2}}
\label{a3}
\end{equation}
This vector potential is plotted in the right panel of Figure \ref{fig7}. Note that $E(t)$ is an even function, while $A(t)$ is an odd function, as in the case of one pair of turning points, shown in Figure \ref{fig1}. The equation for the turning points is a cubic equation in $t^2$, so we obtain three  complex conjugate pairs of turning points, $(t_1(k), t_1^*(k))$,  $(t_2(k), t_2^*(k))$, and $(t_3(k), t_3^*(k))$. We do not write the expressions explicitly, as they are long and not particularly instructive.
These turning points are illustrated in Figure \ref{fig8}, for various values of the longitudinal momentum.
As in the case of a single pair of turning points shown in Figure \ref{fig2}, the point of closest approach of the turning points to the real axis occurs at  $k_\parallel=0$. This is reflected in the momentum spectrum for the three-pair case, shown in Figure 9, which is centered around  $k_\parallel=0$, but in contrast to the momentum spectrum for the two-pair case, shown in Figure 3, which is centered around  a nonzero value of $k_\parallel$.  Also, observe that since the three pairs are almost equidistant from the real axis,  we should expect significant interference effects between the various pairs of turning points, as indeed is seen in Figure \ref{fig9} for both scalar and spinor QED. The momentum spectrum resulting from vector potentials that are odd functions of time [and hence electric fields that are even functions of time] exhibit symmetric oscillations centered around $k_{\parallel}=0$, since for odd gauge fields the phase integrands have the symmetry: $Q_{k_{\parallel}}(t)=Q_{-k_{\parallel}}(-t)$. Therefore equations (\ref{abdot}) and (\ref{abdot2}) remain invariant under the transformations $k_{\parallel}\rightarrow -k_{\parallel}$, and $t\rightarrow -t$. In the WKB framework, this fact is manifest as the symmetry of the turning point distribution under $k_{\parallel}\rightarrow -k_{\parallel}$, as can be seen from Figure  \ref{fig8}.

Figure \ref{fig9} shows a comparison between the approximations (\ref{three-scalar}) and (\ref{three-spinor}) and the exact numerical results, for the particle number as a function of longitudinal momentum. Note the oscillatory behavior of the spectrum, due to the interference terms. Also notice that the interference terms have different signs for scalar and spinor QED, leading to different oscillatory behavior in the longitudinal momentum spectrum.
The agreement between the exact numerical results [solid, blue lines] and the approximate semiclassical expressions [dashed, red lines] is extremely good, both qualitatively and quantitatively.

\section{Pulse Configurations with Flat Envelopes}

A significant advantage of the semiclassical approach is that it provides us with some physical intuition to guide the problem of designing the temporal shape of the electric field $E(t)$ in order to produce a desired momentum specturm. This is an interesting, and difficult, "inverse problem", and in this Section we illustrate the idea with some examples. The treatment of temporally localized electric fields with sub-cycle structure is important within the context of vacuum pair production in that these type of fields represent more realistic pulse configurations with rich structure of momentum spectrum for the produced pairs. Further, from the experimental point of view, the investigation of field parameter dependance of the spectrum might be useful for achieving more  prolific pair production.

The first observation is that interference effects are more likely with an electric field with more temporal structure: the single-bump field exhibits no interference,  while the fields with increasing numbers of maxima and minima tend to increase the level of interference in the momentum spectrum. This is obvious from the scattering picture, but it is not the whole story. We also see that there is a marked difference between cases where $E(t)$ is an even or odd function. Indeed, for a complicated form [with many oscillations and possibly an envelope] of $E(t)$, and hence correspondingly for $A(t)$, the effective scattering potential $-(k_\parallel-A(t))^2$ changes dramatically as a function of $k_\parallel$, and it is not easy to see from the form of this scattering potential when there would be a minimum or maximum of the particle number.
The best indicator comes from looking at the location of the turning points in the complex plane. This also shows us that interference effects will be most pronounced when different sets of turning points are approximately equidistant from the real axis.

We can illustrate these trends with some electric field configurations looking more and more like realistic laser pulses, with sub-cycle structure. In \cite{Hebenstreit:2009km}, the effect of the carrier phase was investigated for fields of a given frequency $\omega$, convolved with a Gaussian envelope function, with a phase offset $\phi$:
\begin{eqnarray}
E(t)=E_0\, \cos(\omega\,t+\phi)\,e^{-t^2/(2 \tau^2)}
\label{gaussian}
\end{eqnarray}
Strong interference effects are seen for the odd field where $\phi=\pi/2$, and we now understand this as due to the interference between two dominant pairs of turning points \cite{dd1}. Now we ask what happens if we change the shape of the field so that more than two pairs of turning points contribute. There should then be stronger interference effects. This can be achieved by "flattening" the envelope function from a Gaussian to a factor $e^{-t^4/\tau^4}$ or $e^{-t^8/\tau^8}$. We show below that this simple change  in the envelope function increases the number of relevant turning point pairs, and correspondingly has a significant effect on the interference terms. We consider such envelope functions both for "cosine-like" and "sine-like" electric fields, corresponding to carrier phases $\phi=0$ and $\phi=\pi/2$, respectively.

\subsection{ Envelope Functions: $\exp[-t^4/\tau^4]$}

\begin{figure}[htb]
\includegraphics[scale=0.7]{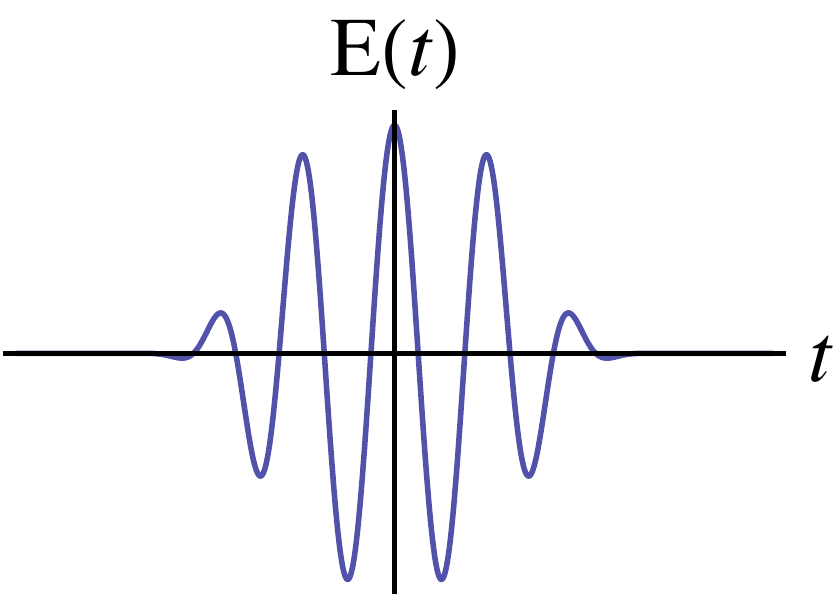}\qquad
\includegraphics[scale=0.7]{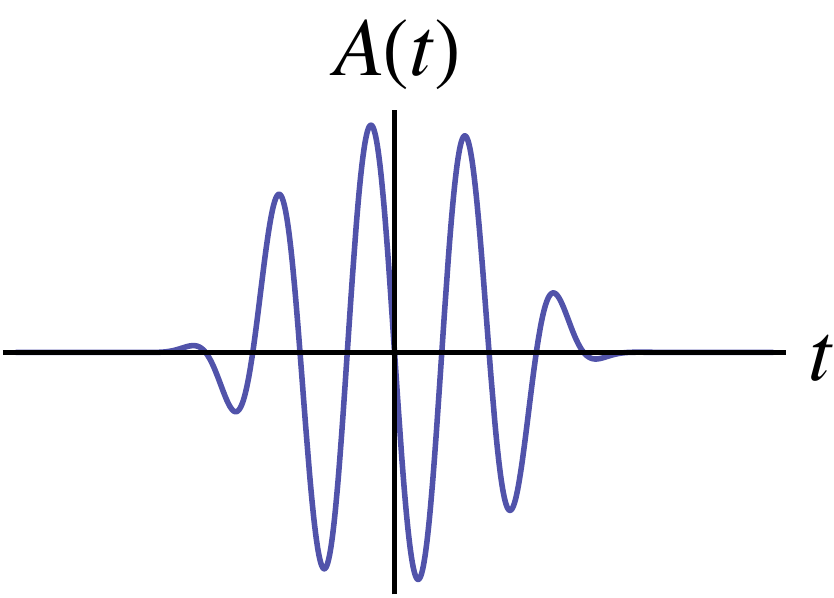}
\caption{The form of the electric field $E(t)$ in (\ref{es1}), and corresponding vector potential $A(t)$ in (\ref{as1}) for two complex conjugate pairs of turning points. $E(t)$ is an even function, while $A(t)$ is an odd function.}
\label{fig10}
\end{figure}

\begin{figure}[htb]
\includegraphics[scale=0.6]{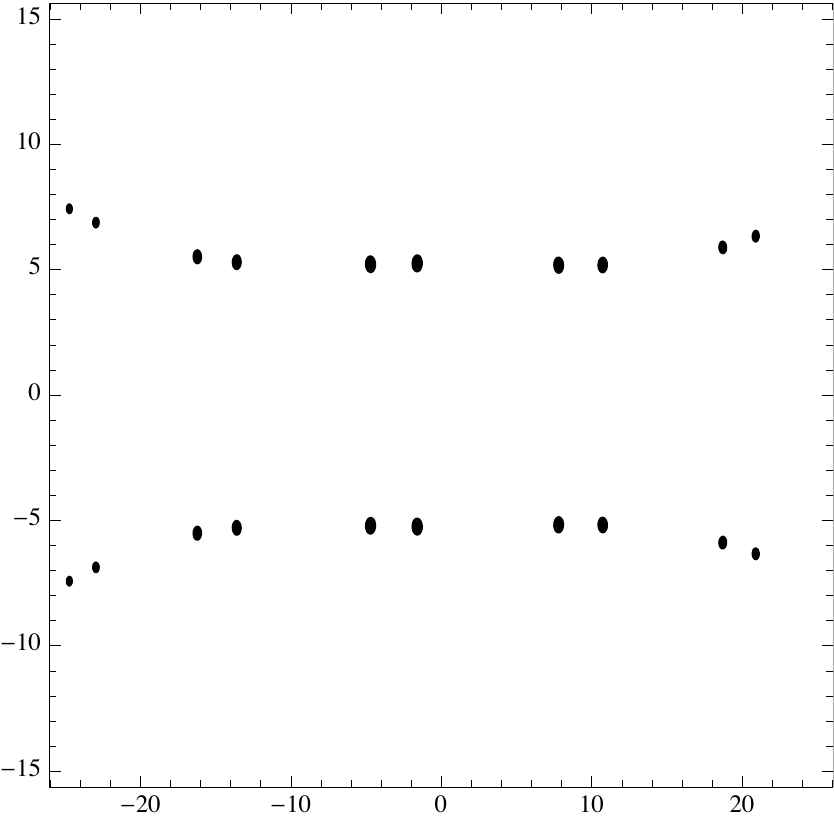}
\includegraphics[scale=0.6]{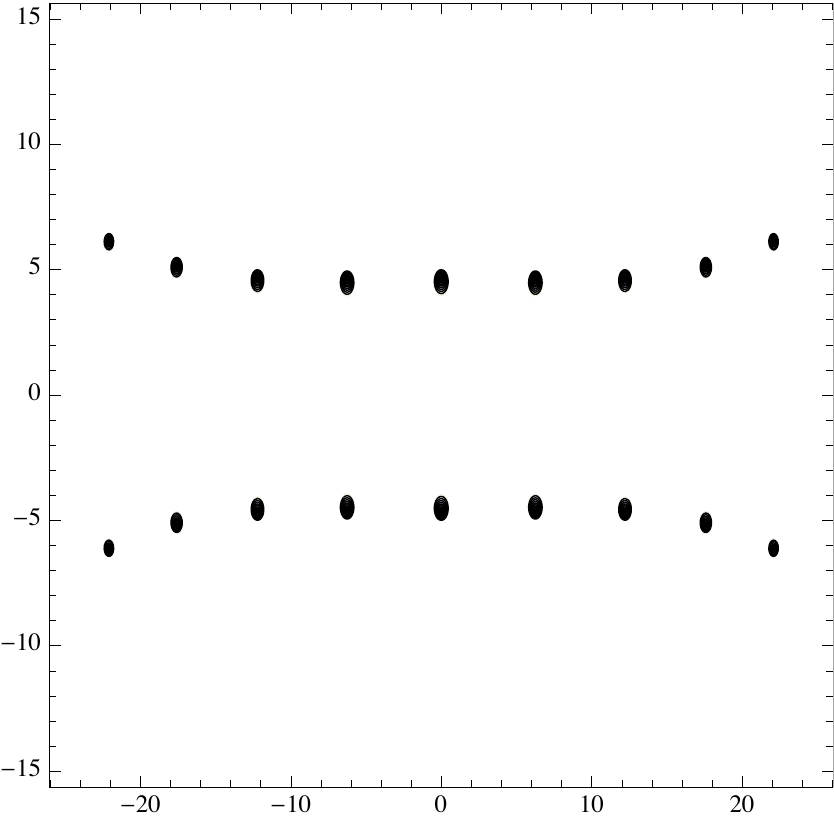}
\includegraphics[scale=0.6]{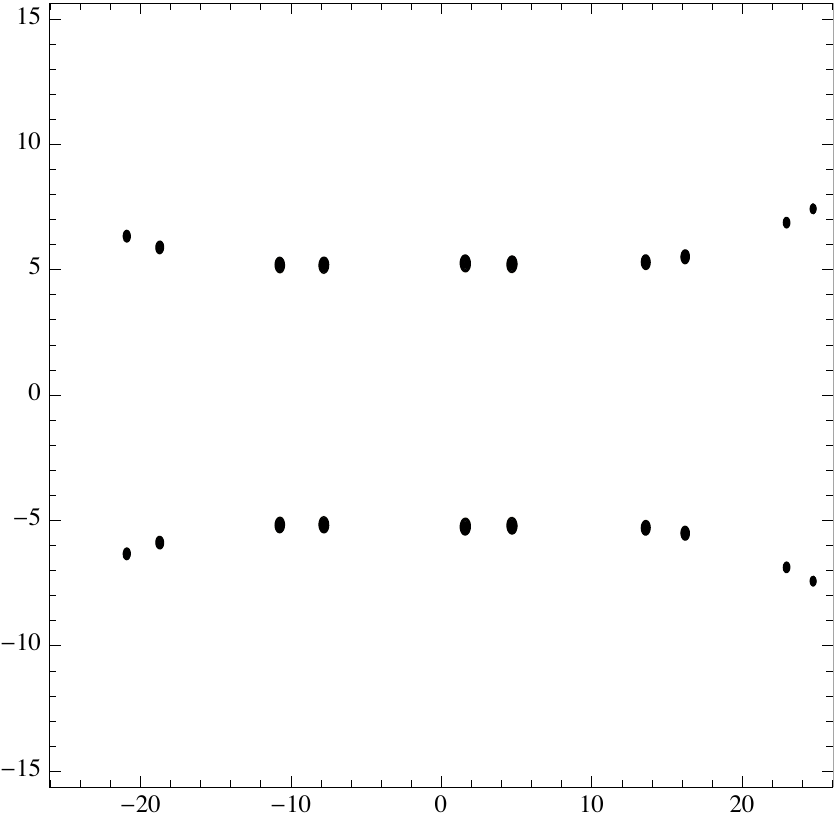}
\caption{The locations of the complex conjugate pair of turning points, in the complex $t$ plane, for three different values of longitudinal momentum. These plots are for the vector potential $A(t)$ in (\ref{as1}), with $E_0=0.1$ and $\omega=0.1$, for longitudinal momentum vales $k_\parallel=-1$ (left), $k_\parallel=0$ (center), and $k_\parallel=1$ (right), in units with $m=1$.}
\label{fig11}
\end{figure}

\begin{figure}[htb]
\includegraphics[scale=0.5]{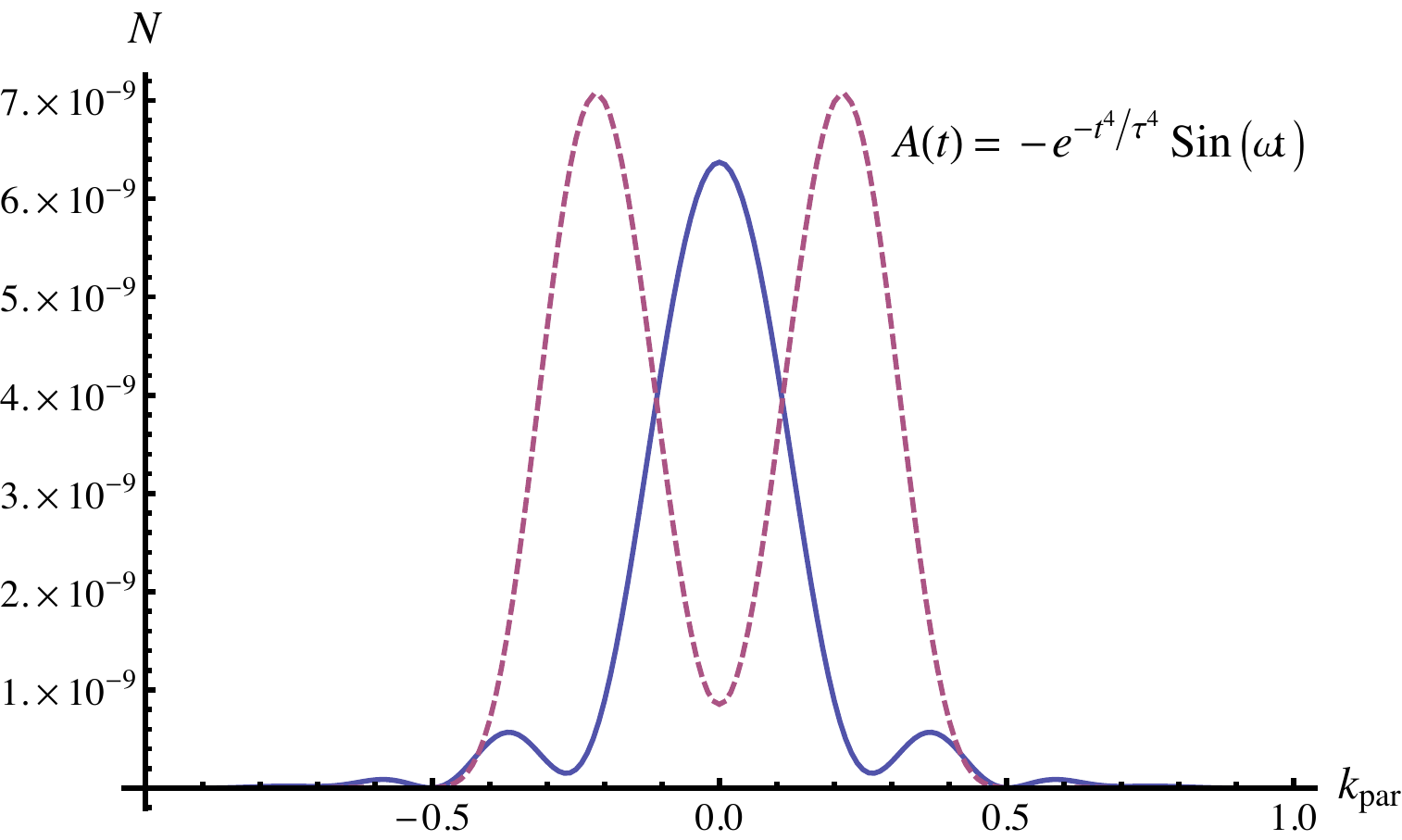}
\caption{The particle numbers for vacuum pair production, as a function of longitudinal momentum, for the electric field (\ref{es1}), with the solid (blue) line showing scalar QED and the dashed (red) line showing spinor QED. The field parameters $E_0$, $\omega$, and $\tau$ were chosen as:\, $E_0=0.1$, $\omega=0.5$, and $\tau=0.05$, in units with $m=1$.}
\label{fig12}
\end{figure}

Consider electric fields with an envelope function $e^{-t^4/\tau^4}$, which is "flatter" than a Gaussian envelope. This leads to more turning points with approximately equal real parts, and therefore to stronger interference effects. Specifically, we first take an electric field temporal profile that is an even function of time
\begin{eqnarray}
E_{\rm even}(t)=\frac{E_0 e^{-\frac{t^4}{\tau ^4}} \left(\tau ^4 \omega  \cos (t \omega )-4
   t^3 \sin (t \omega )\right)}{\tau ^4 \omega }
   \label{es1}
   \end{eqnarray}
which comes from an odd vector potential
\begin{eqnarray}
A(t)=-E_0/\omega\, e^{-t^4/\tau^4}\, \sin(\omega t)
\label{as1}
\end{eqnarray}
The forms of these fields are plotted in Figure \ref{fig10}, and the turning point distribution is sketched in Figure \ref{fig11}. We see that there are more turning point pairs approximately equidistant from the real axis, suggesting stronger interference effects.  The results for the produced particle number, as a function of longitudinal momentum, are shown in Figure \ref{fig12}, for both scalar QED [solid, blue curve] and spinor QED [dashed, red curve]. Notice the single-peak structure for scalar QED, and double-peak structure for spinor QED, a reflection of the opposite sign of interference terms.

\begin{figure}[htb]
\includegraphics[scale=0.7]{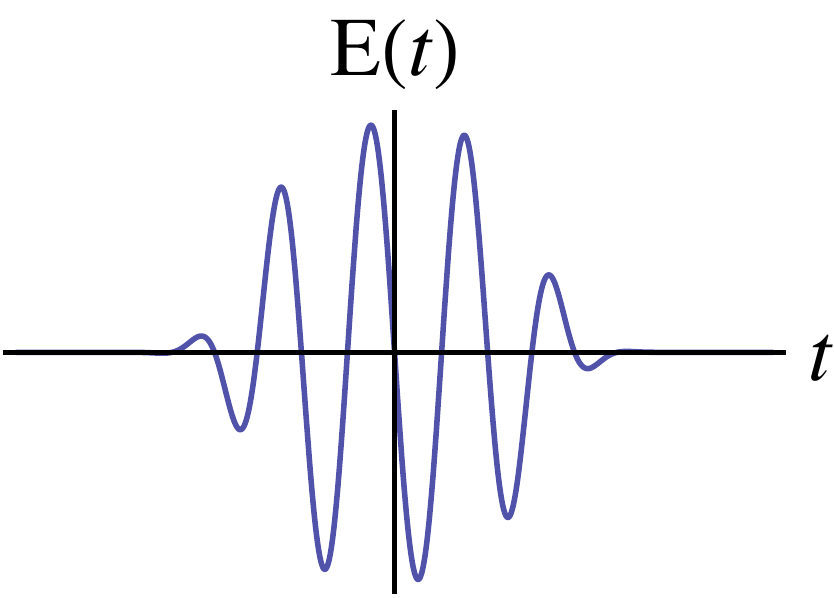}\qquad
\includegraphics[scale=0.7]{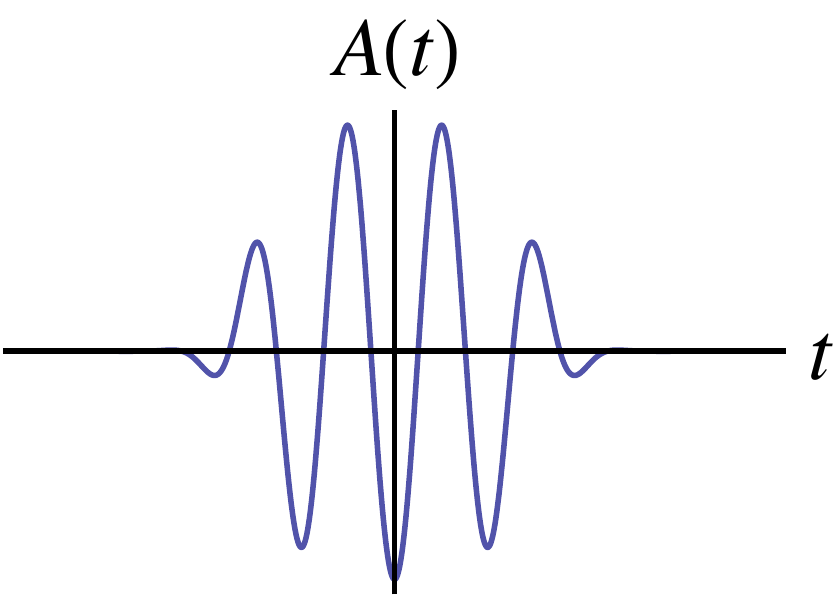}
\caption{The form of the electric field $E(t)$ in (\ref{es2}), and corresponding vector potential $A(t)$ in (\ref{as2}) for two complex conjugate pairs of turning points. $E(t)$ is an odd function, while $A(t)$ is an even function.}
\label{fig13}
\end{figure}

\begin{figure}[htb]
\includegraphics[scale=0.6]{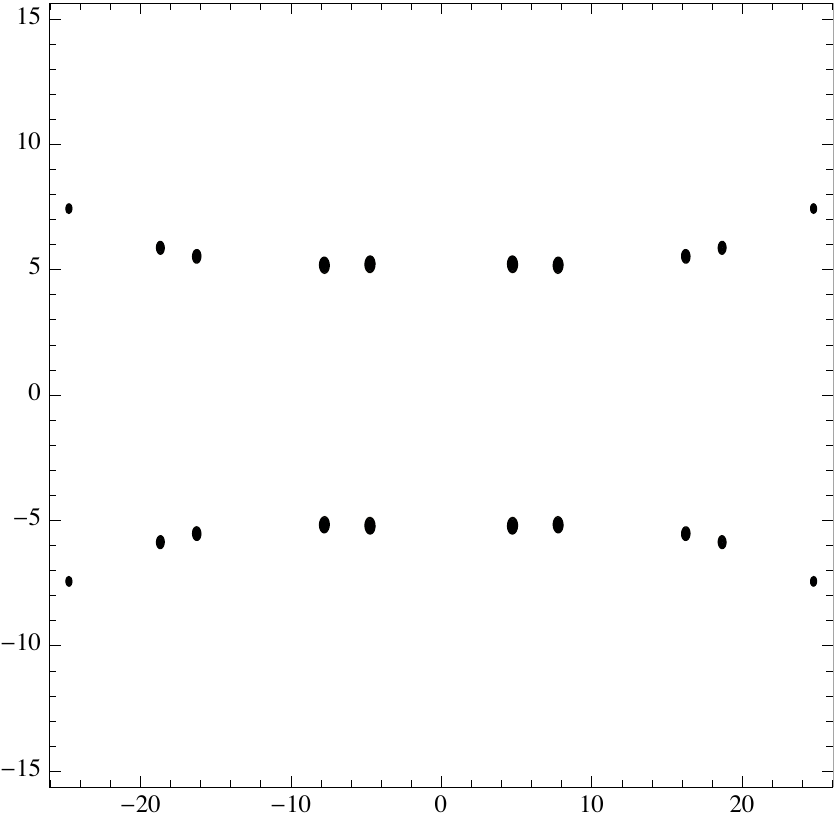}
\includegraphics[scale=0.6]{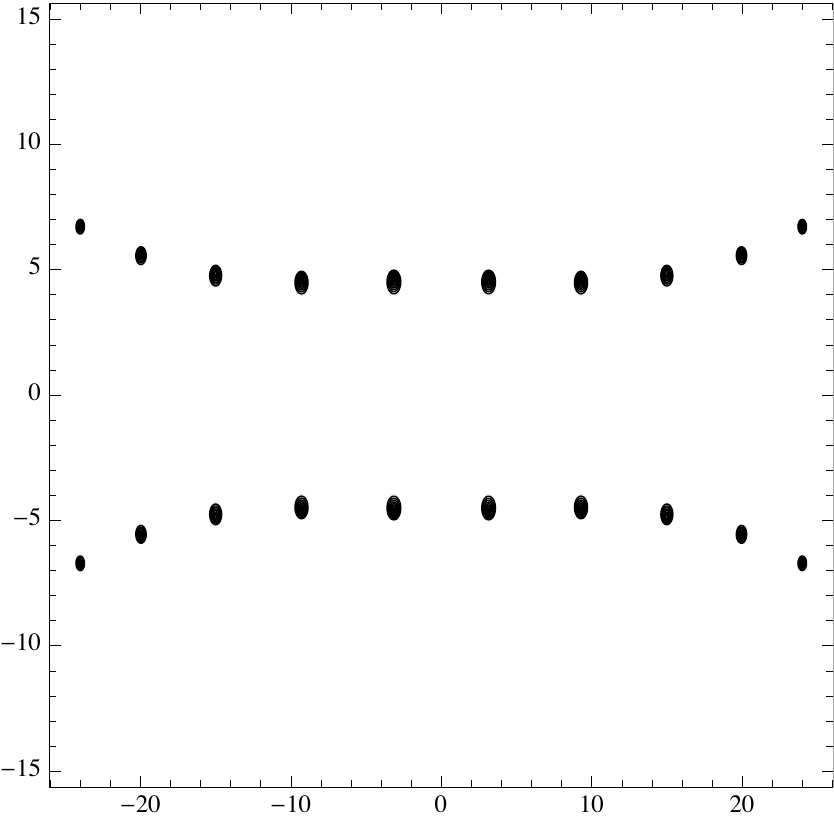}
\includegraphics[scale=0.6]{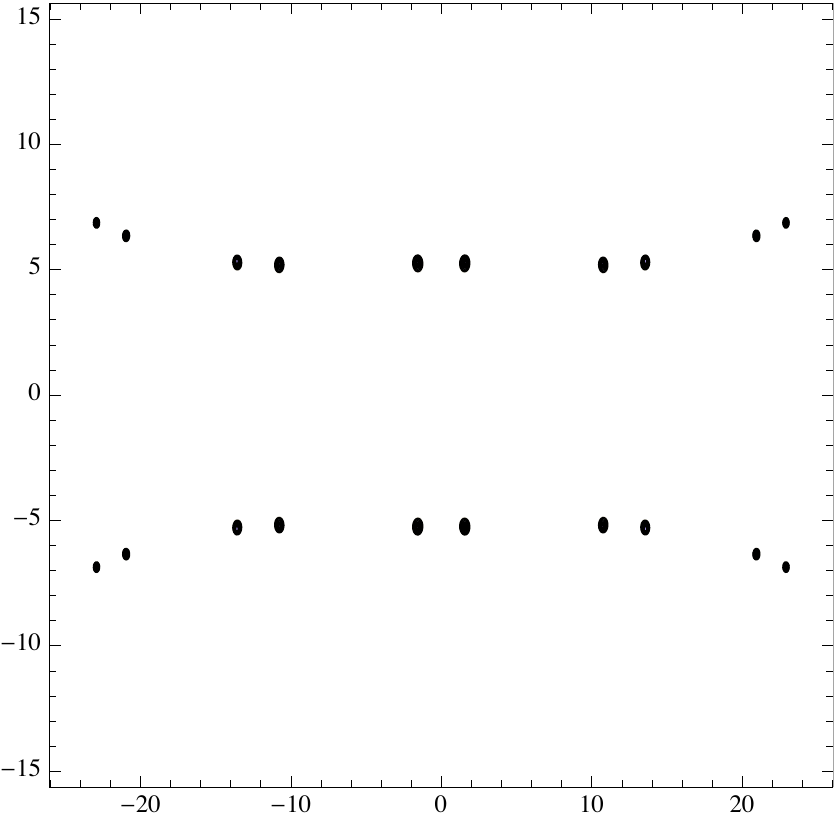}
\caption{The locations of the complex conjugate pair of turning points, in the complex $t$ plane, for three different values of longitudinal momentum. These plots are for the vector potential $A(t)$ in (\ref{as2}), with $E_0=0.1$ and $\omega=0.1$, for longitudinal momentum vales $k_\parallel=-1$ (left), $k_\parallel=0$ (center), and $k_\parallel=1$ (right), in units with $m=1$.}
\label{fig14}
\end{figure}

\begin{figure}[htb]
\includegraphics[scale=0.5]{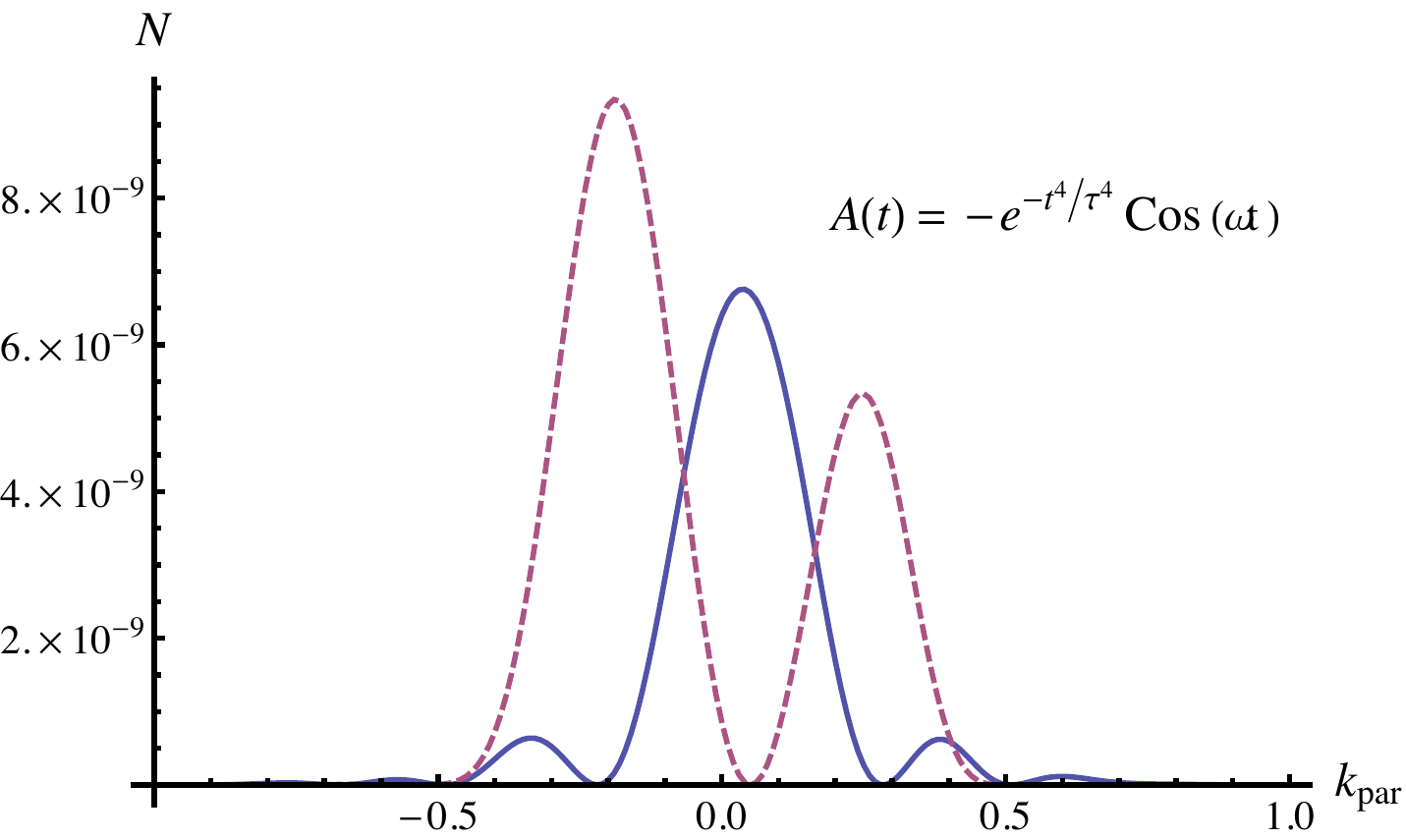}
\caption{The particle numbers for vacuum pair production, as a function of longitudinal momentum, for the electric field (\ref{es2}), with the solid (blue) line showing scalar QED and the dashed (red) line showing spinor QED. The field parameters $E_0$, $\omega$, and $\tau$ were chosen as:\, $E_0=0.1$, $\omega=0.5$, and $\tau=0.05$, in units with $m=1$.}
\label{fig15}
\end{figure}

As a  second example with the same envelope function, we consider an electric field temporal profile that is an odd function of time
\begin{eqnarray}
E_{\rm odd}(t)= -\frac{E_0 e^{-\frac{t^4}{\tau ^4}} \left(4 t^3 \cos (t \omega )+\tau ^4
   \omega  \sin (t \omega )\right)}{\tau ^4 \omega }
   \label{es2}
   \end{eqnarray}
which comes from a vector potential
\begin{eqnarray}
A(t)=-E_0/\omega\, e^{-t^4/\tau^4}\, \cos(\omega t)
\label{as2}
\end{eqnarray}
The forms of these fields are plotted in Figure \ref{fig13}, and the turning point distribution is sketched in Figure \ref{fig14}. We see that there are more turning point pairs approximately equidistant from the real axis, suggesting stronger interference effects. The results for the produced particle number, as a function of longitudinal momentum, are shown in Figure \ref{fig15}, for both scalar QED [solid, blue curve] and spinor QED [dashed, red curve]. Again, notice the single-peak structure for scalar QED, and double-peak structure for spinor QED, but now observe the asymmetry of the spinor spectrum.

\subsection{Envelope Functions: $\exp[-t^8/\tau^8]$}

\begin{figure}[htb]
\includegraphics[scale=0.7]{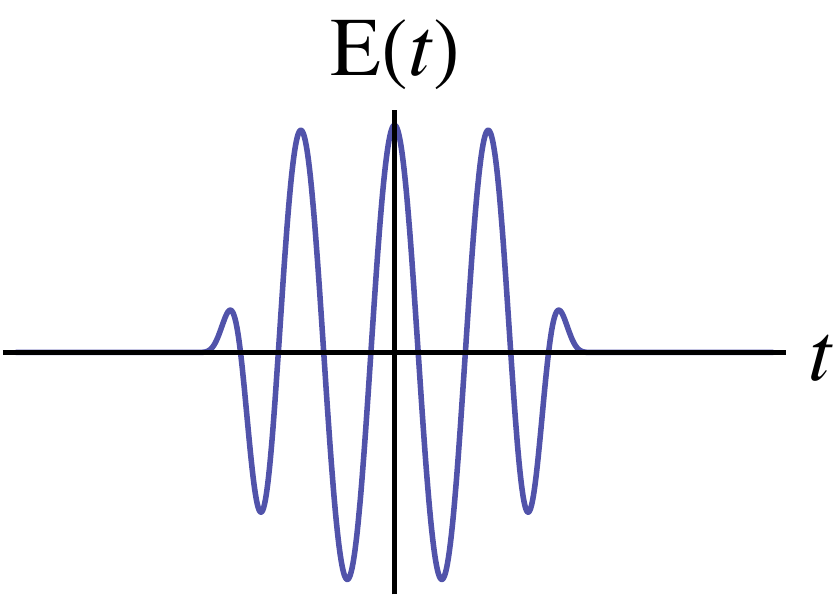}\qquad
\includegraphics[scale=0.7]{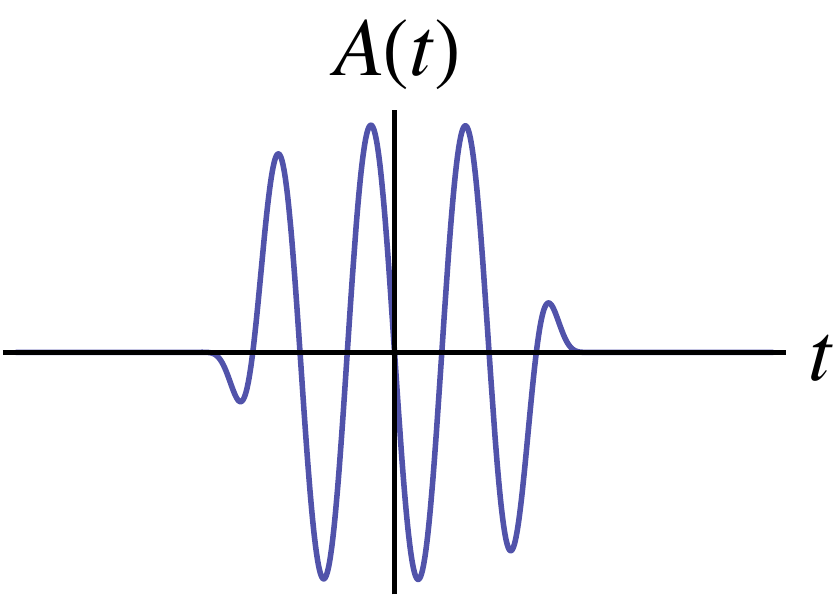}
\caption{The form of the electric field $E(t)$ in (\ref{es3}), and corresponding vector potential $A(t)$ in (\ref{as3}) for two complex conjugate pairs of turning points. $E(t)$ is an even function, while $A(t)$ is an odd function.}
\label{fig16}
\end{figure}

\begin{figure}[htb]
\includegraphics[scale=0.6]{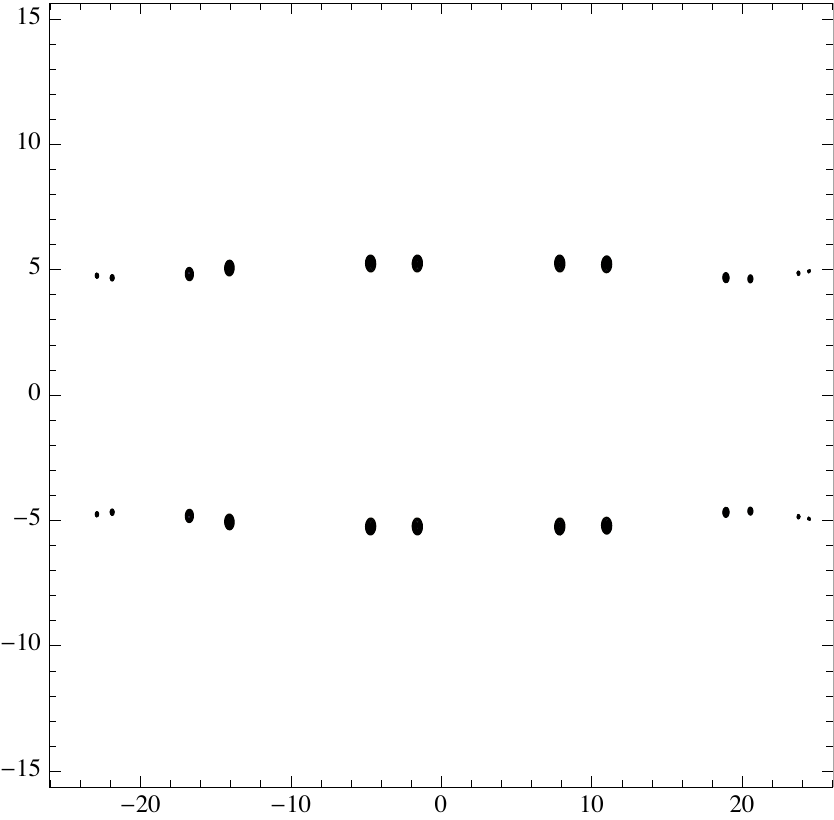}
\includegraphics[scale=0.6]{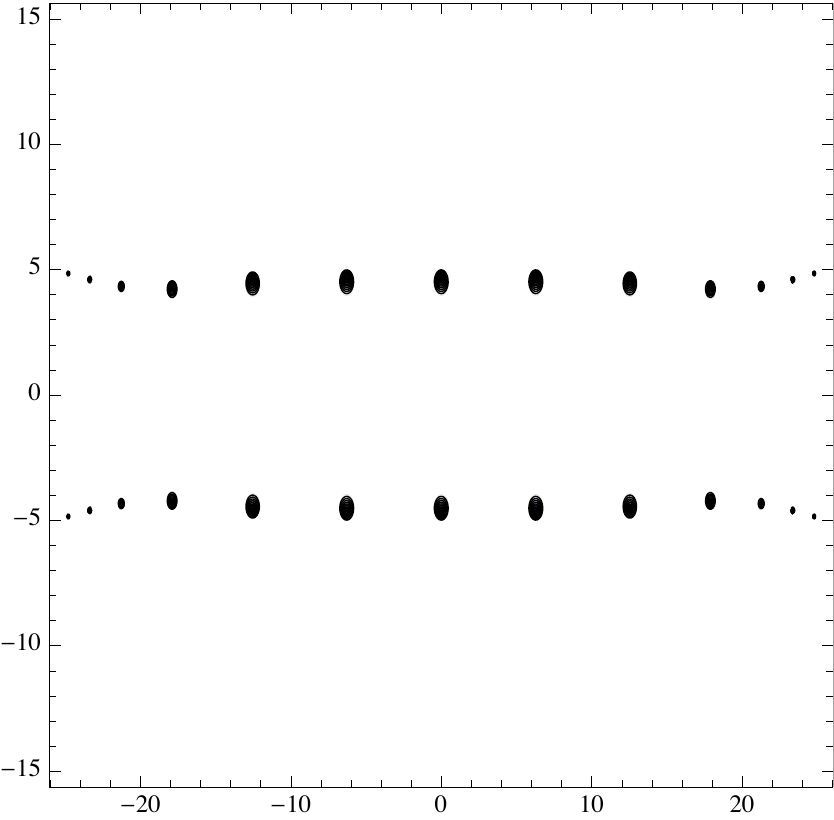}
\includegraphics[scale=0.6]{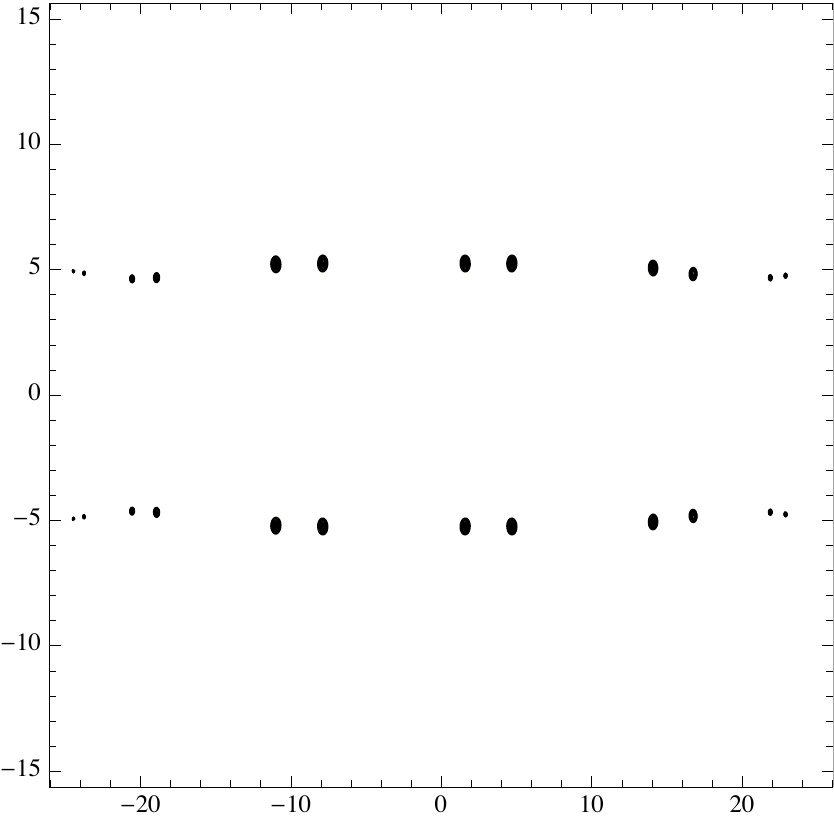}
\caption{The locations of the complex conjugate pair of turning points, in the complex $t$ plane, for three different values of longitudinal momentum. These plots are for the vector potential $A(t)$ in (\ref{as3}), with $E_0=0.1$ and $\omega=0.1$, for longitudinal momentum vales $k_\parallel=-1$ (left), $k_\parallel=0$ (center), and $k_\parallel=1$ (right), in units with $m=1$.}
\label{fig17}
\end{figure}

\begin{figure}[htb]
\includegraphics[scale=0.5]{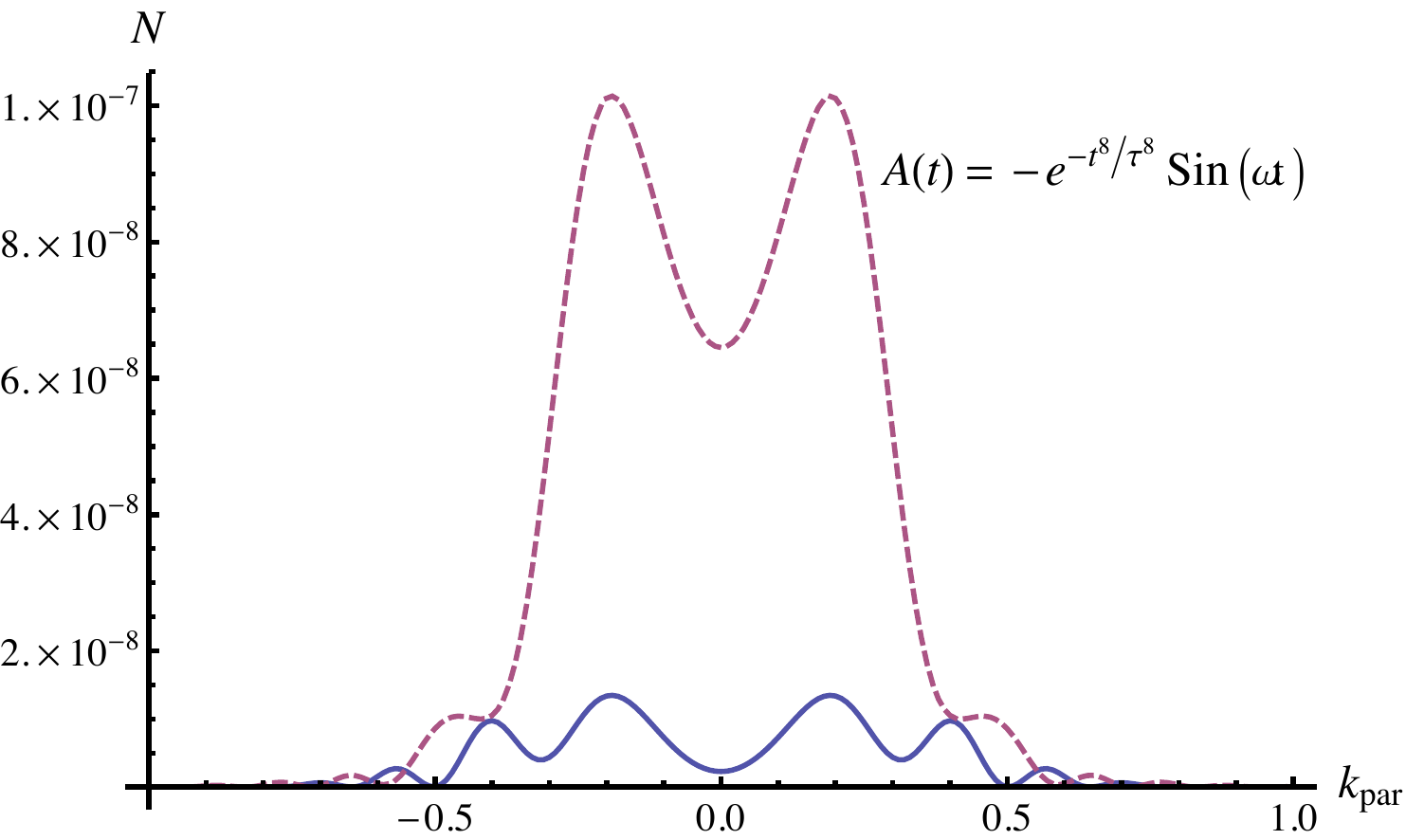}
\caption{The particle numbers for vacuum pair production, as a function of longitudinal momentum, for the electric field (\ref{es3}), with the solid (blue) line showing scalar QED and the dashed (red) line showing spinor QED. The field parameters $E_0$, $\omega$, and $\tau$ were chosen as:\, $E_0=0.1$, $\omega=0.5$, and $\tau=0.05$, in units with $m=1$.}
\label{fig18}
\end{figure}

Now consider fields with an even flatter envelope function: $e^{-t^8/\tau^8}$. This leads to even more turning points with approximately equal real parts, and therefore to even stronger interference effects. Specifically, we first take an electric field temporal profile that is an even function of time
\begin{eqnarray}
E_{\rm even}(t)= \frac{E_0 e^{-\frac{t^8}{\tau ^8}} \left(\tau ^8 \omega  \cos (t \omega )-8
   t^7 \sin (t \omega )\right)}{\tau ^8 \omega }
   \label{es3}
   \end{eqnarray}
which comes from an odd vector potential
\begin{eqnarray}
A(t)=-E_0/\omega\, e^{-t^8/\tau^8}\, \sin(\omega t)
\label{as3}
\end{eqnarray}
The forms of these fields are plotted in Figure \ref{fig16}, and the turning point distribution is sketched in Figure \ref{fig17}. We see that there are even more turning point pairs approximately equidistant from the real axis, suggesting stronger interference effects.  The results for the produced particle number, as a function of longitudinal momentum, are shown in Figure \ref{fig18}, for both scalar QED [solid, blue curve] and spinor QED [dashed, red curve]. Notice the very different forms of the momentum spectra, and in particular notice that the peak values for spinor QED are almost an order of magnitude greater than for scalar QED.

\begin{figure}[htb]
\includegraphics[scale=0.7]{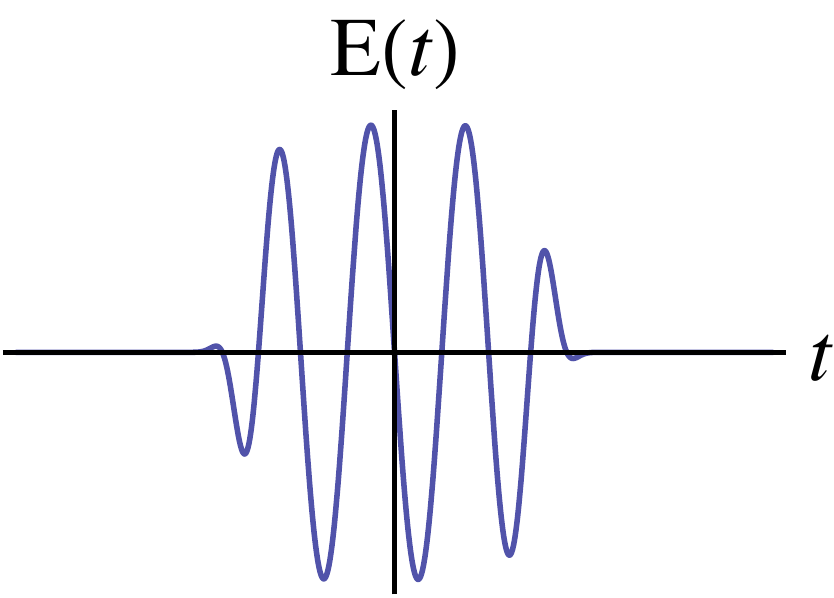}\qquad
\includegraphics[scale=0.7]{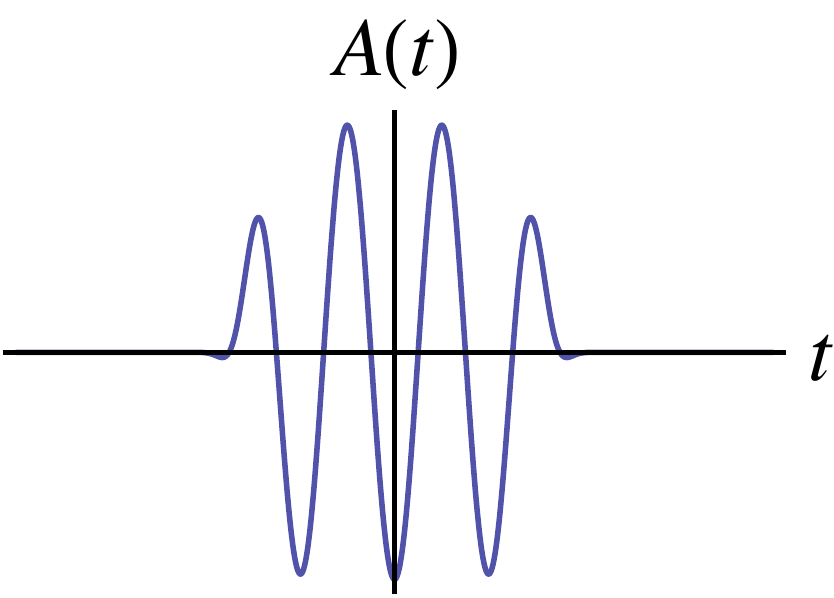}
\caption{The form of the electric field $E(t)$ in (\ref{es4}), and corresponding vector potential $A(t)$ in (\ref{as4}) for two complex conjugate pairs of turning points. $E(t)$ is an odd function, while $A(t)$ is an even function.}
\label{fig19}
\end{figure}

\begin{figure}[htb]
\includegraphics[scale=0.6]{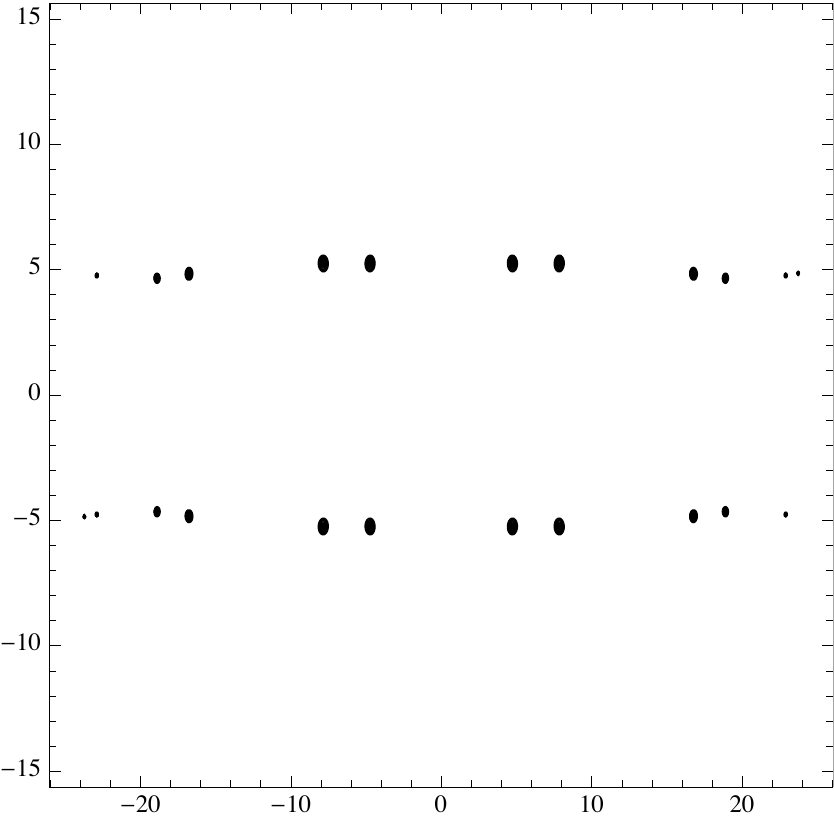}
\includegraphics[scale=0.6]{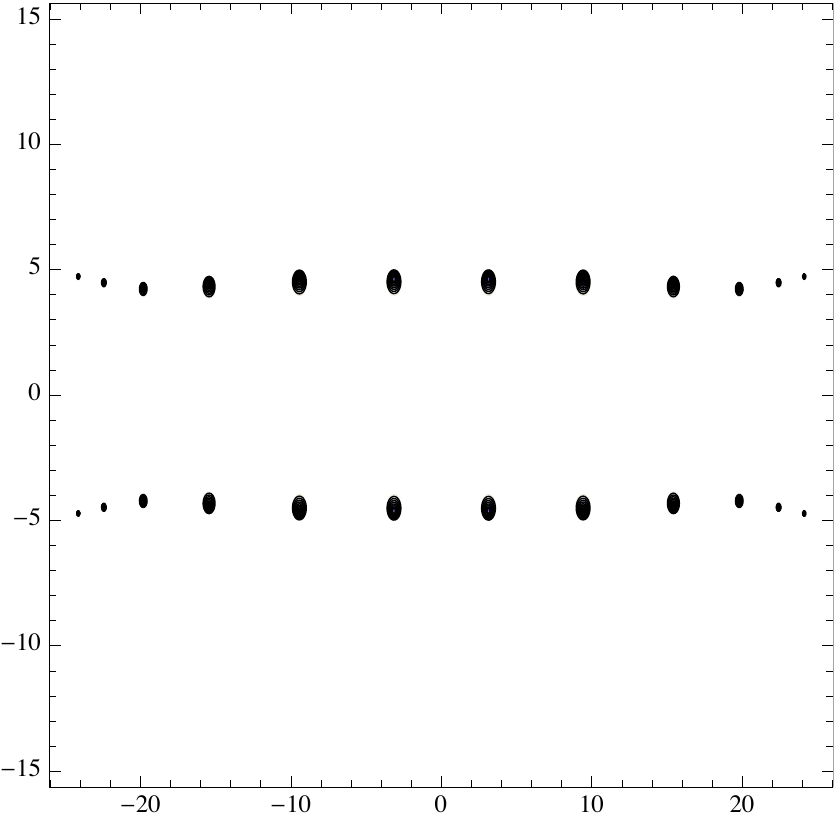}
\includegraphics[scale=0.6]{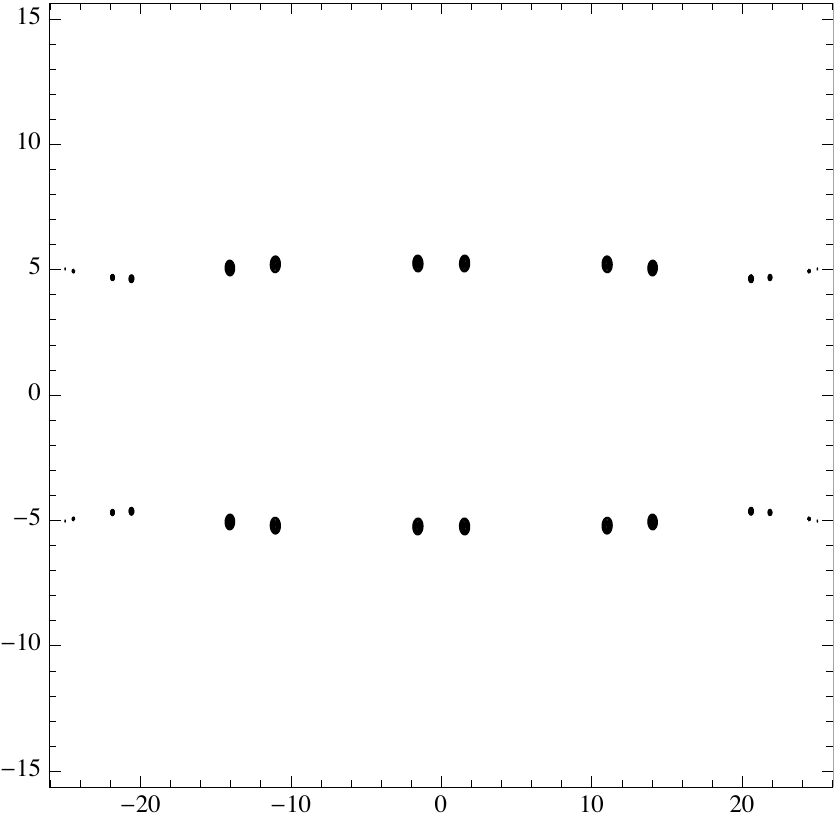}
\caption{The locations of the complex conjugate pair of turning points, in the complex $t$ plane, for three different values of longitudinal momentum. These plots are for the vector potential $A(t)$ in (\ref{as4}), with $E_0=0.1$ and $\omega=0.1$, for longitudinal momentum vales $k_\parallel=-1$ (left), $k_\parallel=0$ (center), and $k_\parallel=1$ (right), in units with $m=1$.}
\label{fig20}
\end{figure}

\begin{figure}[htb]
\includegraphics[scale=0.5]{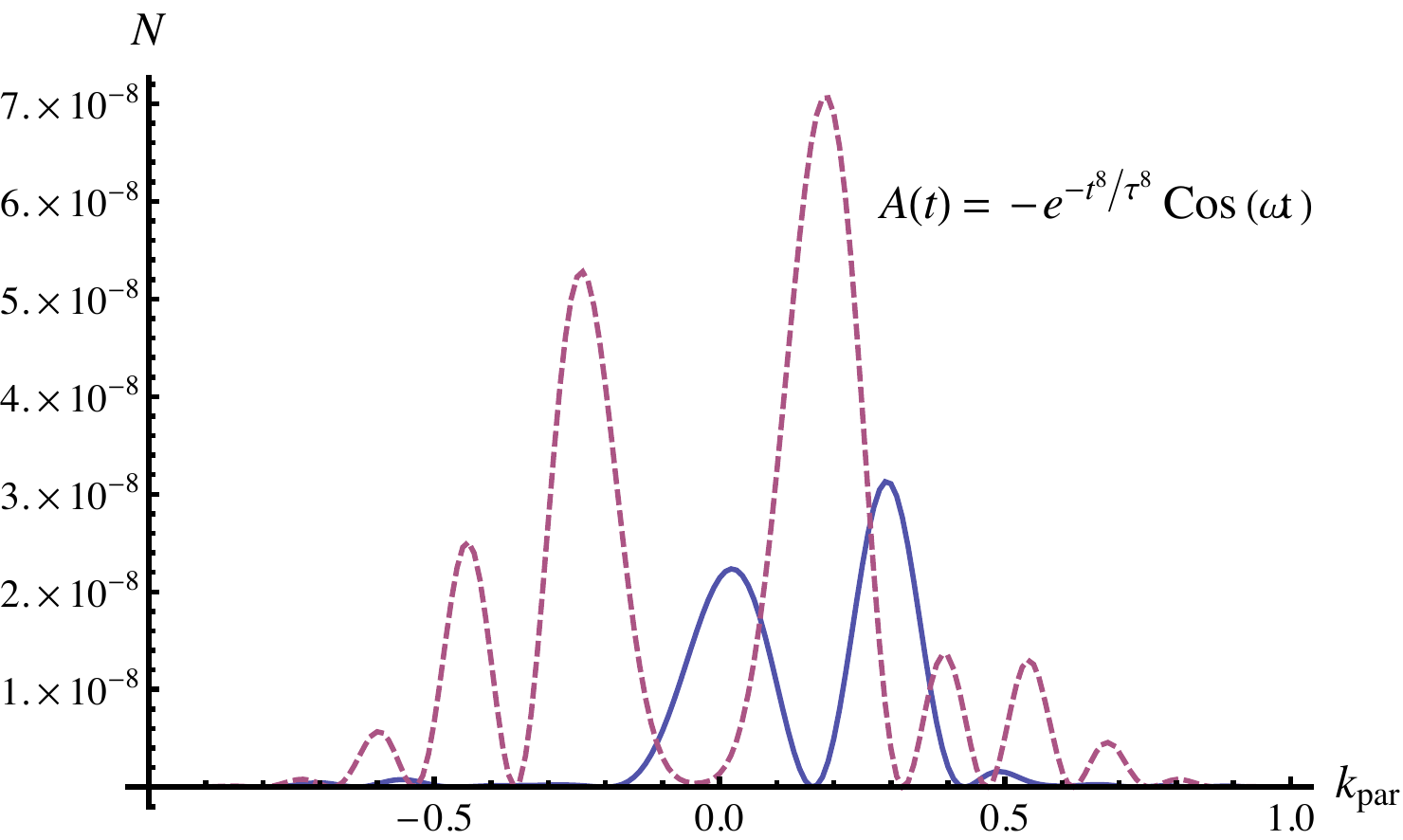}
\caption{The particle numbers for vacuum pair production, as a function of longitudinal momentum, for the electric field (\ref{es4}), with the solid (blue) line showing scalar QED and the dashed (red) line showing spinor QED. The field parameters $E_0$, $\omega$, and $\tau$ were chosen as:\, $E_0=0.1$, $\omega=0.5$, and $\tau=0.05$, in units with $m=1$.}
\label{fig21}
\end{figure}

As a  second example with the same envelope function, we consider an electric field temporal profile that is an odd function of time
\begin{eqnarray}
E_{\rm odd}(t)=-\frac{E_0 e^{-\frac{t^8}{\tau ^8}} \left(8 t^7 \cos (t \omega )+\tau ^8
   \omega  \sin (t \omega )\right)}{\tau ^8 \omega }
   \label{es4}
   \end{eqnarray}
which comes from a vector potential
\begin{eqnarray}
A(t)=-E_0/\omega\, e^{-t^8/\tau^8}\, \cos(\omega t)
\label{as4}
\end{eqnarray}
The forms of these fields are plotted in Figure \ref{fig19}, and the turning point distribution is sketched in Figure \ref{fig20}. Again, we see that there are even more turning point pairs approximately equidistant from the real axis, suggesting stronger interference effects.  The results for the produced particle number, as a function of longitudinal momentum, are shown in Figure \ref{fig21}, for both scalar QED [solid, blue curve] and spinor QED [dashed, red curve]. Notice the very different form of the spectra, and note that again the spinor QED peaks are noticeably higher than those for scalar QED.

These examples clearly show that the flatter the envelope function, the stronger the interference effects, and with such a  large number of turning points participating, there can be  large differences between the pair production for spinor and scalar QED. For example, in Figure \ref{fig18} we see almost an order of magnitude difference between the spinor and scalar QED peak particle numbers. Contrast this with the earlier examples, where even though the interference effects have different signs, they do not conspire to increase the overall magnitude of the peak values.

\section{Conclusions}

In this paper we have investigated interference effects in the longitudinal momentum spectrum for particles produced from vacuum by a linearly polarized electric field that is spatially uniform but time dependent.
The interference is due to the interaction between multiple semiclassical turning points, and becomes important when the temporal profiles have subcycle structure, as is true for more realistic laser pulse fields than just the well-studied single-bump fields like $E(t)=E_0\,{\rm sech}^2(\omega\,t)$.
We have given a simple new approximate formulas, (\ref{app-scalar}) and (\ref{app-spinor}), for the number of produced particles, as a function of longitudinal momentum, for both scalar and spinor QED, for an arbitrary number of turning points, extending the result of \cite{dd1} for the interference between two distinct turning points.
As expected, the interference terms have different signs depending on the particle statistics. We have confirmed that these approximate expressions agree very well with the exact results, obtained by numerical integration of the Riccati form of the corresponding scattering problem, for electric fields having precisely one, two, and three complex  conjugate pairs of semiclassical turning points. The approximate expressions provide important physical intuition that may be used to guide the shaping of the temporal profile of electric field pulses in order to obtain a particular momentum spectrum. In particular, we have shown that flattening the temporal envelope function leads to stronger interference effects, since more turning points interfere, and tends to increase the particle number for spinor QED relative to scalar QED. We hope that this semiclassical approach may be useful in guiding the design of planned laser experiments in order to observe this elusive non-perturbative Heisenberg-Schwinger effect for the first time. For example, the recent numerical results of Orthaber et al \cite{Orthaber:2011cm} concerning the momentum spectrum of vacuum particle production for the dynamically assisted Schwinger mechanism \cite{Schutzhold:2008pz}, in which a strong enhancement is seen when a weak but rapidly varying field is superimposed on a stronger but slower field, can  be understood semiclassically in terms of the appearance of new saddle points that arise due to the additional weak field.
In addition, such time-dependent tunneling problems appear in many other contexts \cite{KeskiVakkuri:1996gn}, in particle and nuclear physics, condensed matter physics, atomic physics, chemical physics, and gravitational physics, and we anticipate that the simplicity of these results may prove useful in these other areas also. Finally, the semiclassical perspective in terms of interfering saddle points may prove useful in the search for a computationally effective formalism that also incorporates spatial inhomogeneities of the laser pulses, for example using worldline instantons \cite{wli} or Wigner function methods \cite{Hebenstreit:2010vz,Hebenstreit:2010cc}.

\bigskip

We  acknowledge support from the DOE  grant DE-FG02-92ER40716.

\end{document}